\documentclass[acmtog, nonacm]{acmart}

\acmJournal{TOG}

\usepackage{overpic}
\usepackage{arydshln}
\usepackage{amsmath}
\usepackage{graphics}
\usepackage{placeins} 
\usepackage{makecell} 
\usepackage{wrapfig}
\usepackage{amssymb}
\usepackage{multirow}
\usepackage{enumerate}
\usepackage{color}

\newcommand{\XR}[1]{{ #1}}

\newcommand{\site}{\mathbf{x}}
\newcommand{\decay}{\tau}
\newcommand{\rve}{\mathbf{e}}
\newcommand{\under}[1]{\underline{\textbf{#1}}}

\setcopyright{acmlicensed}
\acmJournal{TOG}
\acmYear{2024} \acmArticle{1} 
\acmVolume{44} \acmNumber{4} \acmMonth{8} \acmPrice{15.00}
\acmDOI{10.1145/xxxxxxx}

\begin{document}

\title{CWF: Consolidating Weak Features in High-quality Mesh Simplification}


\author{Rui Xu}
\authornote{\ \  Equal contribution. \\
$\dagger$ Corresponding authors.}
\affiliation{\institution{Shandong University} 
\country{China}}
\email{xrvitd@163.com}

\author{Longdu Liu}
\authornotemark[1]
\affiliation{  \institution{Shandong University}
\country{China}}\email{liulongdu@163.com}

\author{Ningna Wang}
\affiliation{  \institution{The University of Texas at Dallas}
\country{USA}}
\email{ningna.wang@utdallas.edu}

\author{Shuangmin Chen}
\affiliation{  \institution{Qingdao University of Science and Technology}
\country{China}}\email{csmqq@163.com}

\author{Shiqing Xin}
\authornotemark[2]
\affiliation{  \institution{Shandong University}
\country{China}}\email{xinshiqing@sdu.edu.cn}

\author{Xiaohu Guo}
\affiliation{  \institution{The University of Texas at Dallas}
\country{USA}}\email{xguo@utdallas.edu}

\author{Zichun Zhong}
\affiliation{  \institution{Wayne State University}
\country{USA}}\email{zichunzhong@wayne.edu}

\author{Taku Komura}
\affiliation{  \institution{The University of Hong Kong}
\country{China}}\email{taku@cs.hku.hk}

\author{Wenping Wang}
\affiliation{  \institution{Texas A\&M University}
\country{USA}}\email{wenping@tamu.edu}

\author{Changhe Tu} 
\authornotemark[2]
\affiliation{  \institution{Shandong University}
\country{China}}
\email{chtu@sdu.edu.cn}


\begin{abstract}
In mesh simplification, common requirements like accuracy, triangle quality, and feature alignment are often considered as a trade-off. 
Existing algorithms concentrate on just one or a few specific aspects of these requirements. For example, the well-known Quadric Error Metrics (QEM) approach~\cite{garland1997surface} prioritizes accuracy and can preserve strong feature lines/points as well, but falls short in ensuring high triangle quality and may degrade weak features that are not as distinctive as strong ones.
In this paper, we propose a smooth functional that simultaneously considers all of these requirements. 
The functional comprises a normal anisotropy term and a Centroidal Voronoi Tessellation~(CVT)~\cite{du1999centroidal} energy term, with the variables being a set of movable points lying on the surface. The former inherits the spirit of QEM but operates in a continuous setting, while the latter encourages even point distribution, allowing various surface metrics. We further introduce a decaying weight to automatically balance the two terms.
We selected 100 CAD models from the ABC dataset~\cite{koch2019abc}, along with 21 organic models, to compare the existing mesh simplification algorithms with ours.
Experimental results reveal an important observation: the introduction of a decaying weight effectively reduces the conflict between the two terms and enables the alignment of weak features. This distinctive feature sets our approach apart from most existing mesh simplification methods and demonstrates significant potential in shape understanding. Please refer to the teaser figure for illustration.

\end{abstract}



\begin{teaserfigure}
  \centering
  \vspace{-2mm}
  \includegraphics[width=\textwidth]{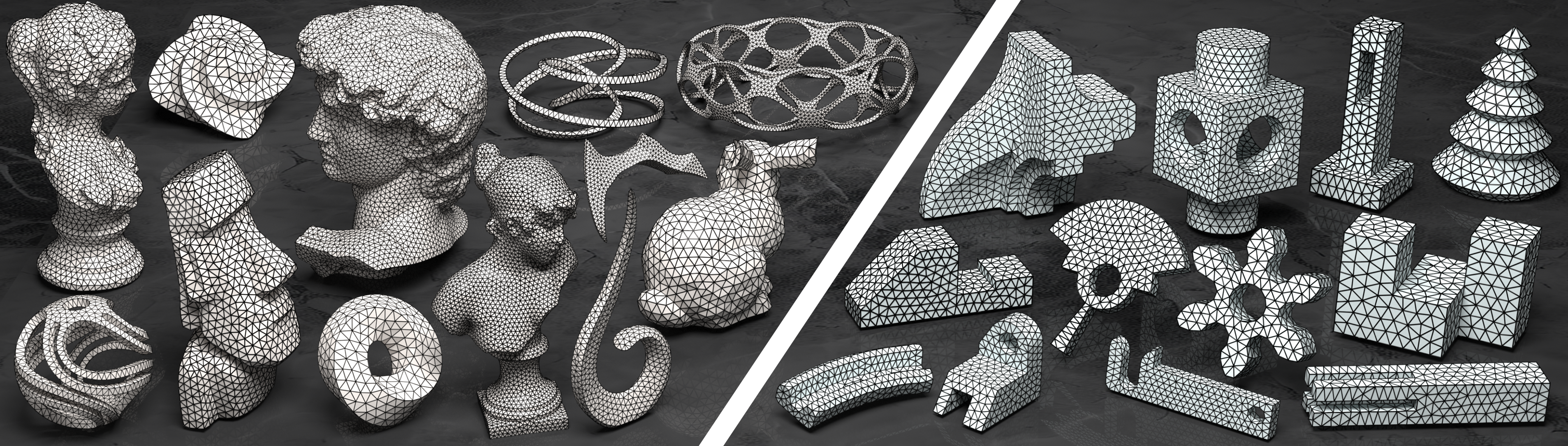}
  \vspace{-1mm}
  \vspace{-6mm}
  \caption{
  In this paper, we propose a functional for aligning both strong and weak feature lines during mesh simplification while maintaining high-quality triangulations. The left side shows organic models, and the right side displays CAD models. It's important to mention that the original versions of the organic models are heavily smoothed.
  }
  \label{fig:teaser}
\end{teaserfigure}


\maketitle
\newcommand{\rspace}{\mathbb{R}}
\newcommand{\vorcell}{\Omega^{vor}}
\newcommand{\anypoint}{\mathbf{x}}
\newcommand{\sample}{\mathbf{p}}
\newcommand{\query}{\mathbf{q}}
\newcommand{\cross}{\mathcal{A}}
\newcommand{\area}{a}
\newcommand{\normal}{\mathbf{n}}
\newcommand{\numV}{M}
\newcommand{\numP}{N}
\newcommand{\ie}{\textit{i.e., }}
\newcommand{\eg}{\textit{e.g., }}

\vspace{-10pt}

\section{Introduction}
\label{sec:Introduction}

Mesh simplification aims to reduce complexity to balance the need of visual fidelity with computational efficiency. It has been proved useful in various applications. First, a simplified mesh can be considered as a more compact representation~\cite{dou2020top, li2021feature} w.r.t the original mesh, which can facilitate real-time transmission over networks~\cite{ko20023d,cabiddu2015large}, and rendering in mobile applications or AR/VR environments~\cite{bahirat2018designing,fu2022easyvrmodeling}. Second, a simplified mesh serves as a proxy of the original shape, requiring fewer computational resources and enhancing speed in interactive applications~\cite{zhang2023hessian} and simulations.

Common requirements in the field of mesh simplification include accuracy~\cite{lescoat2020spectral,liu2023surface}, triangle quality~\cite{du1999centroidal,CVTwang2018isotropic,abdelkader2020vorocrust}, and feature alignment~\cite{garland1997surface,levy2010p,chen2023robust}. Accuracy is typically measured based on the difference between the simplified and original versions. Good triangle quality not only leads to more accurate results in physics simulations but also increases the stability of numerical computations~\cite{dou2022coverage, wang2022computing, wang2024coverage,wang2024mattopo}. Feature alignment helps provide crucial visual clues for effective shape recognition and involves aligning the mesh edges with strong and weak feature lines. However, weak features, less distinctive than strong ones, are challenging to preserve or even consolidate during mesh simplification.


In past research, requirements such as accuracy, triangle quality, and feature alignment were often considered trade-offs. Existing algorithms, including Quadric Error Metrics (QEM)~\cite{garland1997surface} and Centroidal Voronoi Tessellation (CVT)~\cite{du1999centroidal}, focus on a few specific aspects but struggle to achieve a good balance. For instance, QEM, the most popular approach for mesh simplification, involves a sequence of basic operations such as edge contraction. Despite its advantages in accuracy and alignment with strong features, it suffers from the degradation of weak features. In contrast, CVT focuses on maintaining an even point distribution by optimizing a set of movable points, yet it falls short in aligning with any feature lines, whether strong or weak.

In this paper, we propose a smooth functional that concurrently addresses the aforementioned requirements. This functional includes a normal anisotropy term and a CVT energy term, applied to a set of movable points on the surface. The normal anisotropy term, drawing inspiration from QEM, operates in a continuous setting, while the CVT term promotes an even distribution of points, accommodating various surface metrics. We also introduce a decaying weight to automatically balance these two terms. Optimization typically concludes within tens of iterations, with each iteration decomposing the surface into Voronoi cells. Considering that existing tools for computing Restricted Voronoi Diagrams (RVDs) cannot ensure the ``one site, one region'' property on thin-plate models, we generate an inward counterpart for each movable point and develop a simple yet effective technique to facilitate the computation of RVD on thin-plate models.

We tested our functional on 100 CAD models from the ABC dataset~\cite{koch2019abc} and 21 organic models. The experimental results not only confirm that the normal anisotropy term effectively preserves accuracy and aligns strong features, while the CVT term promotes even point distribution, but they also demonstrate that the introduction of a decaying weight effectively reduces the conflict between the two terms, achieving a good balance among multiple requirements. \XR{This advantageous property makes it useful for the lightweight representation of CAD models.} Additionally, it can be observed that our algorithm is particularly effective in consolidating weak features of organic shapes during mesh simplification.

Our contributions are three-fold:
\vspace{-1mm}
\begin{enumerate}
    \item We propose a functional \XR{that unifies the requirements of accuracy, triangle quality, and feature alignment, effectively achieving a good balance among these key aspects.}    
    \item We introduce a decaying weight to gradually reduce the impact of the CVT energy, thereby naturally coordinating the two terms.
    \item We develop a simple yet effective technique to facilitate the computation of the Restricted Voronoi Diagram (RVD) on thin-plate models.
\end{enumerate}

\begin{figure*}[!t]
    \centering
    \begin{overpic}[width=0.99\linewidth]{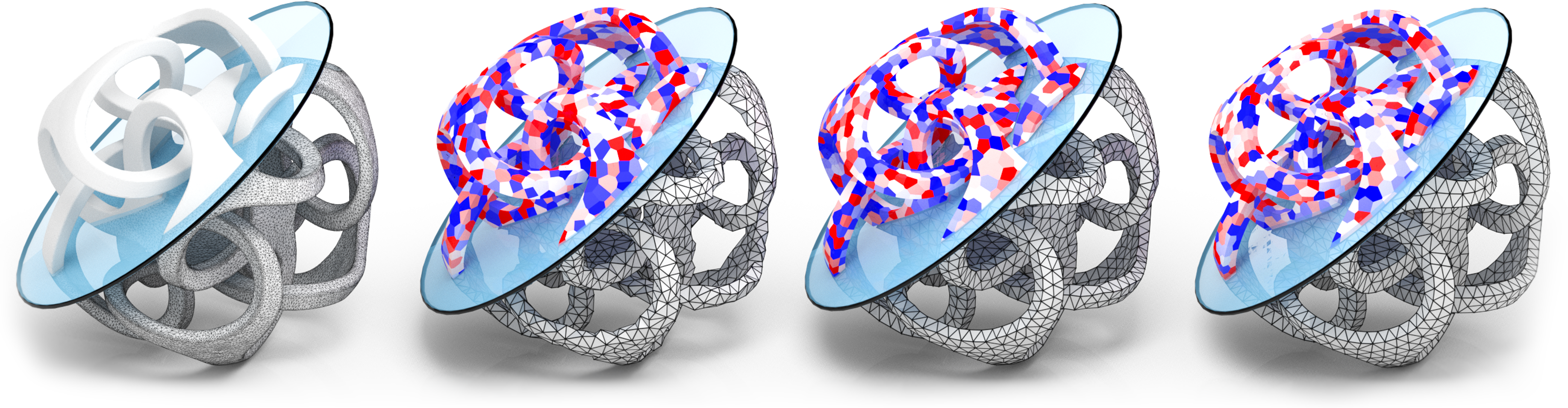}
    \put(7,  0){\textbf{(a) Input Surface}}
    \put(32, 0){\textbf{(b) First Iteration}}
    \put(57, 0){\textbf{(c) 10th Iteration}}
    \put(81, 0){\textbf{(d) 50th Iteration}}
    \end{overpic}
    \vspace{-8pt}
    \caption{
    We present a functional that integrates the demands of accuracy, triangle quality, and feature alignment. The impact of the CVT energy gradually diminishes, thanks to the decaying weight, achieving a harmonious balance between the two terms. It is noted that a Restricted Voronoi Diagram (RVD) must be computed during each iteration. In this example, a total of 50 iterations activated our termination condition.}
    \label{fig:Pipeline}
    \vspace{-8pt}
\end{figure*}
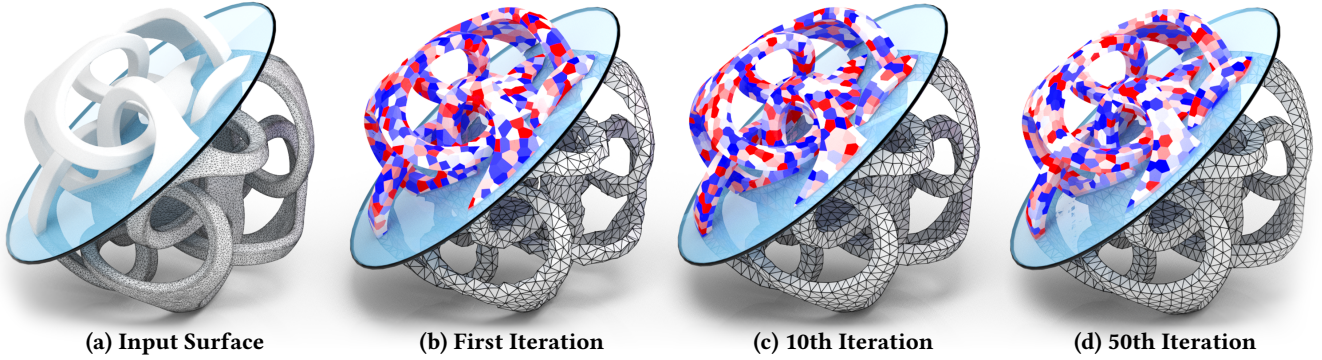

\vspace{-4mm}
\section{Related Work}

Triangle mesh surfaces have become increasingly popular and widely used. Given that dense triangulations can lead to high computational costs, mesh simplification is essential for balancing computational efficiency with an acceptable level of visual or geometric fidelity across various applications. Key considerations typically include accuracy, triangle quality, and feature alignment. In the following, we will review related works on mesh simplification.

\subsection{CVT and Its Variants}
By minimizing the Centroidal Voronoi Tessellation (CVT) energy~\cite{du1999centroidal}, one can achieve an even distribution of movable points on a surface. There is a substantial body of literature on using CVT to generate high-quality mesh surfaces~\cite{valette2004approximated,valette2008generic}. A notable implementation combines Restricted Voronoi Diagrams (RVD) and L-BFGS optimization to accomplish high-quality meshing, as proposed in~\cite{yan2009isotropic}. Additionally, numerous variants of CVT~\cite{sun2011obtuse,edwards2013kCVT,CVTyan2015non,du2018field,CVTwang2018isotropic} have been developed to enhance triangulations in specific scenarios. However, most of these variants struggle with feature alignment.

In the work of LpCVT~\cite{levy2010p}, normal anisotropy is incorporated into the objective function to facilitate feature alignment during remeshing. However, the balancing weight between the two terms requires careful case-by-case adjustment according to the specific input shape. Typically, the normal anisotropy term approaches zero for CAD models, but remains significant for organic models. This variability leads to LpCVT's inconsistency in preserving both strong and weak features.

\subsection{Accuracy Guided Mesh Simplification}
Local decimation schemes are widely favored in mesh simplification for their ease of implementation and near-linear scalability~\cite{hoppe1993mesh}. To minimize accuracy loss, \citet{garland1997surface} introduced a fast, high-quality edge collapse-based simplification algorithm using the Quadric Error Metric (QEM). This algorithm is proficient at predicting target positions during edge contraction into a vertex, utilizing the normal vectors of the incident triangles. Our tests confirm that QEM offers an accuracy advantage and can align with strong features. Due to its popularity, various QEM variants have been developed. For example, \citet{ozaki2015out} suggested dividing the original surface into multiple patches and performing QEM-guided simplification on each to enhance parallelism and ensure algorithmic feasibility. \citet{wei2010feature} expanded the classic QEM into a constrained high-dimensional space with specific considerations for singular scenarios. Additional improvements to QEM are discussed in~\cite{panchal2022feature,liu2023surface, chen2023robust}.

The primary goal of preserving accuracy is to minimize the Hausdorff distance between the original and simplified surfaces, ensuring geometric fidelity~\cite{ma2012novel}. Alternatives include defining an isotropic density function or an anisotropic metric to guide mesh simplification. For instance, \citet{du2005anisotropic} integrated the traditional Lloyd-CVT method with anisotropic Riemannian metrics, naming it ACVT. \citet{zhong2013particle} achieved anisotropic meshing by mapping the anisotropic space to a higher-dimensional isotropic space. \citet{xu2019anisotropic} worked directly with anisotropic meshes generated from existing remeshing algorithms, focusing on eliminating obtuse angles. Moreover, curvature has been utilized to generate adaptive and isotropic mesh results~\cite{su2019curvature,lv2022adaptively}. However, these algorithms primarily improve accuracy rather than focusing on feature alignment.

\subsection{Feature Preserving Mesh Simplification}
Feature preservation is a critical aspect of mesh simplification. Several approaches~\cite{xie2012surface, yan2013gap, yan2014blue, zhong2013particle} begin by pre-detecting features and subsequently conducting re-meshing while retaining these features. For instance, VoroCrust~\cite{abdelkader2020vorocrust} suggests explicitly placing points symmetrically along pre-detected feature lines. While pre-detecting features seems reasonable, it becomes impractical for organic models where identifying weak features is challenging compared to strong features.

A more promising approach involves ensuring that movable points naturally align with underlying features. Most existing algorithms leverage the inherent property of QEM, which naturally captures strong features. For instance, \citet{valette2008generic} used QEM to guide the generation of Voronoi vertices, ensuring feature alignment. Similarly, \citet{chiang2011robust} proposed mesh quality improvement through quadratic error-based mesh relaxation. Additionally, \citet{gao2013feature} extended optimal Delaunay triangulation (ODT) to surface meshes by solving a quadratic optimization problem. Although recent works~\cite{panchal2022feature, zhao2023variational,xu2022rfeps} have utilized QEM's ability to preserve sharp features, it must be noted that weak features remain a challenge for QEM.

Various approaches utilize an anisotropy field or curvature information to optimize point placement. \citet{levy2010p} employ an anisotropy field to guide point placement without the need for tagging. In contrast, \citet{jakob2015instant} introduce a field-aligned meshing method for isotropic triangular/quad-dominant remeshing, incorporating local smoothing and control over sharp features. Additionally, \citet{cai2016surface} propose a novel PCA-based energy for face-based clustering, resulting in an anisotropic mesh surface. However, these methods face challenges in achieving high-quality isotropic triangulations and struggle to preserve weak features effectively.

\section{Formulation}
In this paper, we introduce a unified functional designed to meet three essential requirements: accuracy, triangle quality, and feature alignment. Utilizing a set of movable points $\{\mathbf{x}_i\}_{i=1}^N$, as inputs, our approach alternates between decomposing the surface into distinct regions based on proximity and optimizing point distribution. This effectively decouples point placement from their connections, a characteristic reminiscent of the principles underlying CVT. 
\subsection{Objective Function}
\XR{Our objective function is constructed as follows:}
\begin{equation}
    E(\{\mathbf{x}_i\}_{i=1}^N) = \lambda_{\text{NA}} E_{\text{NA}} + \lambda_{\text{CVT}} E_{\text{CVT}},
\label{eq:ourObj}
\end{equation}
where $E_{\text{NA}}$ and $E_{\text{CVT}}$ respectively denote the normal anisotropic term and the CVT term. The normal anisotropic term~\cite{levy2010p} is defined as:
\begin{equation}
E_{\text{NA}}=\sum_i\int_{\Omega_i}\left((\mathbf{x}-\mathbf{x}_i)^\text{T} \mathbf{n}_\mathbf{x}\right)^2 \text{d}s = \sum_i\int_{\Omega_i}(\mathbf{x}-\mathbf{x}_i)^\text{T} \mathbf{n}_\mathbf{x} \mathbf{n}_\mathbf{x}^\text{T} (\mathbf{x}-\mathbf{x}_i) \text{d}s,
\end{equation}
where $\{\Omega_i\}_{i=1}^N$ defines the surface decomposition, commonly computed as a Restricted Voronoi Diagram (RVD). The CVT term is expressed as:
\begin{equation}
E_{\text{CVT}}=\sum_{i=1}^N\int_{\Omega_i}(\mathbf{x}-\mathbf{x}_i)^\text{T}M_{\text{CVT}} (\mathbf{x}-\mathbf{x}_i)  \text{d}s.
\end{equation}
Typical choices for $M_{\text{CVT}}$ include:
\begin{enumerate}
    \item An identity matrix, defining the Euclidean-line distance.
    \item An isotropic matrix, encoding a density function on the surface.
    \item An anisotropic matrix, utilizing the anisotropic tensor to capture and represent directional variations.
\end{enumerate}

In summary, our objective is represented as:
\begin{equation}
    E(\{\mathbf{x}_i\}_{i=1}^N) = \sum_{i=1}^N\int_{\Omega_i}(\mathbf{x}-\mathbf{x}_i)^\text{T}M (\mathbf{x}-\mathbf{x}_i)  \text{d}s,
\end{equation}
where the kernel matrix $M$ is defined as:
\begin{equation}
M = \lambda_{\text{NA}} M_{\text{NA}} +  \lambda_{\text{CVT}} M_{\text{CVT}}.
\label{eq:M}
\end{equation}

\subsection{Links to Existing Approaches}
\label{subsec:Links}
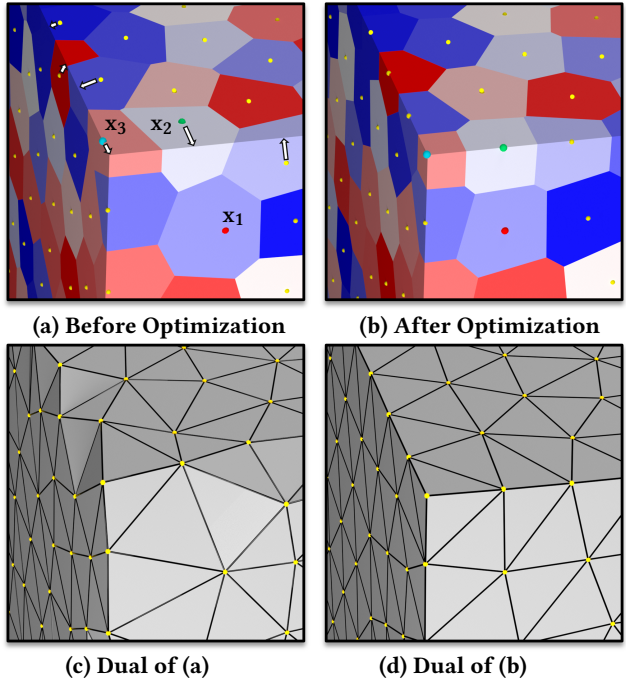
\begin{figure}[!htp]
    \centering
    \begin{overpic}[width=0.98\linewidth]{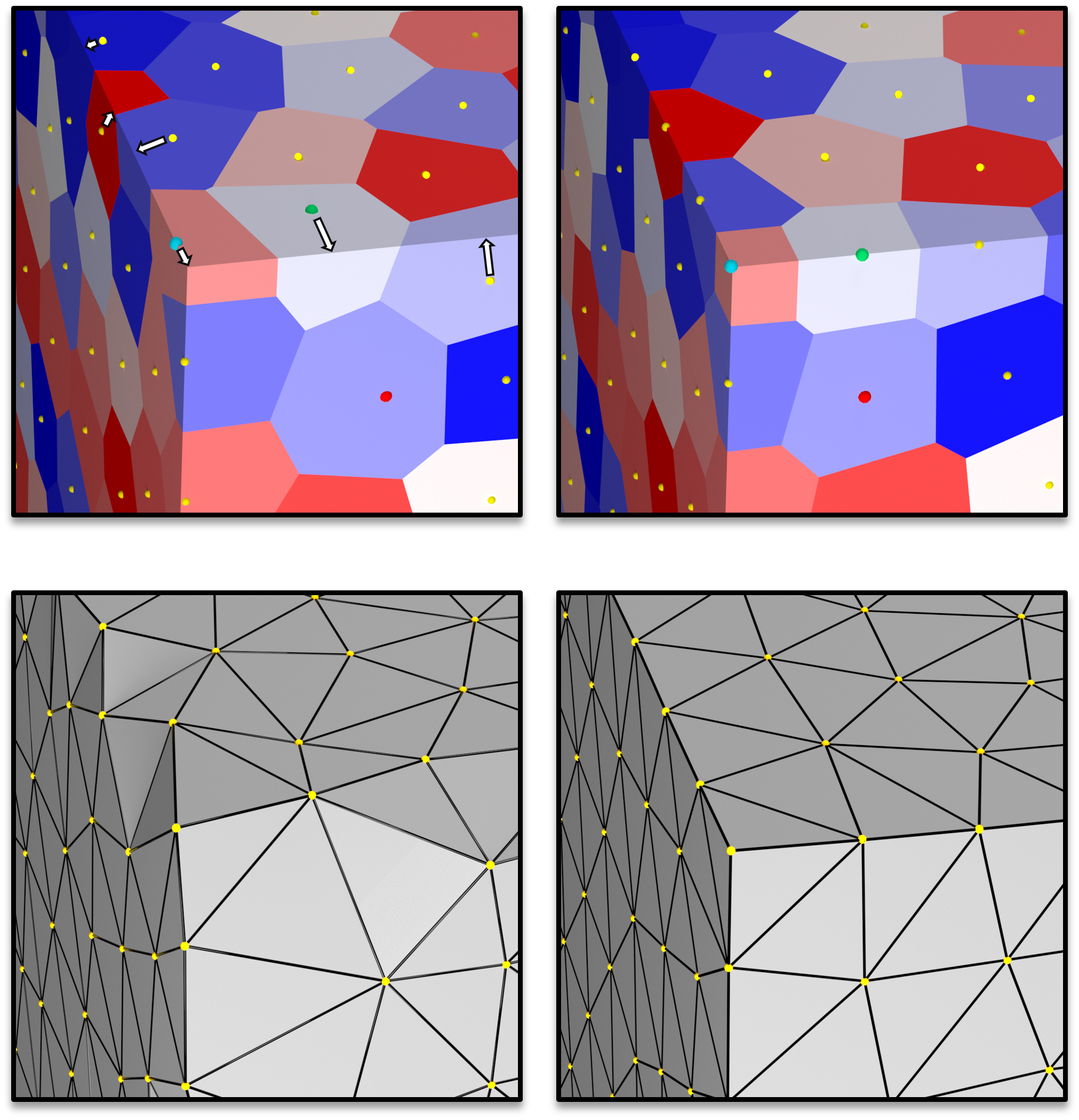}
    \put(5, 49){\textbf{(a) Before Optimization}}
    \put(55, 49){\textbf{(b) After Optimization}}
    \put(10, -3){\textbf{(c) Dual of (a)}}
    \put(58, -3){\textbf{(d) Dual of (b)}}
    \put(23, 80){\textcolor{black}{$\mathbf{x_2}$}}
    \put(34, 66){\textcolor{black}{$\mathbf{x_1}$}}
    \put(16, 80){\textcolor{black}{$\mathbf{x_3}$}} 
    \end{overpic}
    \vspace{-1pt}
    \caption{
    The RVD after optimizing the positions by minimizing our objective function, points near the features are naturally drawn toward the nearby points/lines. In this figure, $\mathbf{x}_1$~\XR{(red)} controls a planar region and remains static, $\mathbf{x}_2$~\XR{(green)} dominates an area across the feature line and is shifted to the feature line, and $\mathbf{x}_3$~\XR{(blue)} governs a corner point and is relocated to the corner.}
    \label{fig:RVD}
    \vspace{-4mm}
\end{figure}

\paragraph{Normal Anisotropy}
As mentioned in~\cite{levy2010p}, the normal anisotropy matrix $\mathbf{n}_\mathbf{x} \mathbf{n}_\mathbf{x}^\text{T}$ quantifies the extent to which the vector $\mathbf{x}-\mathbf{x}_i$ is orthogonal to the normal vector at $\mathbf{x}$, effectively positioning $\mathbf{x}_i$ with respect to all the points in its dominant region~$\Omega_i$.

It is evident that the normal anisotropy term inherits the essence of QEM~\cite{garland1997surface}. The key difference is that while QEM assesses the quadratic error in relation to triangles incident to a mesh edge and its two endpoints, our approach evaluates it within the region of $\Omega_i$. Fig.~\ref{fig:RVD}~(a) illustrates the initial positions and the surface decomposition. The dual of the surface decomposition, shown in Fig.~\ref{fig:RVD}~(c), generates a triangle mesh, but it does not align well with feature points or lines. After optimizing the positions by minimizing our objective function, as demonstrated in Fig.~\ref{fig:RVD}~(b, d), points near features are naturally pulled towards nearby points or lines. To be more specific, $\mathbf{x}_1$ (red) controls a planar region and remains static, $\mathbf{x}_2$ (green) dominates an area across the feature line and shifts towards it, and $\mathbf{x}_3$ (blue) governs a corner point and relocates to that corner.

\XR{It is worth noting that the normal anisotropy term can also achieve its minimum when Voronoi cell boundaries coincide with the model’s feature lines. However, such Voronoi diagrams must have degree-4 vertices, which are not stable configurations. Given a random initial point placement, it is unlikely to terminate at such an unstable configuration, as explained in Fig. 2 of~\cite{liu2009centroidal}. We provide an example of this initialization in Section~\ref{sec:Experimental}.}

\paragraph{Our Considerations}
As discussed in Section~\ref{sec:Introduction}, the essential requirements for mesh simplification include accuracy, triangle quality, and feature alignment. It is clear that the normal anisotropy term primarily preserves accuracy and enforces feature alignment, while the CVT term tends to enhance triangle quality. Achieving a harmonious balance between these aspects is not straightforward. Our analysis identifies two main challenges.

First, the magnitudes of the normal anisotropy term and the CVT term can differ significantly. For example, the normal anisotropy term can be reduced to zero in a standard cube model, but the CVT term cannot be completely eliminated. Conversely, in the case of an organic model, such as the Bunny model, the normal anisotropy term cannot be reduced to zero. This discrepancy underscores the need for a more effective mechanism to balance these terms.

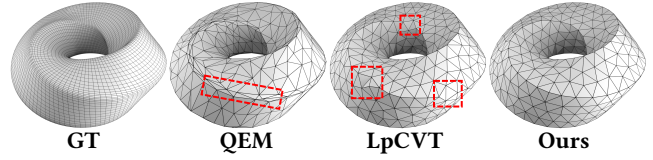
\begin{figure}[!tp]
    \centering
    \begin{overpic}[width=0.99\linewidth]{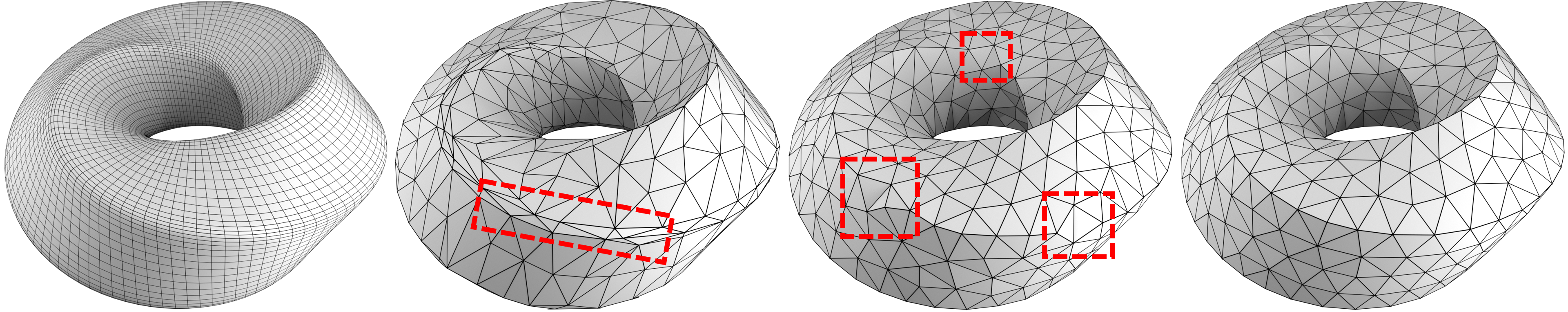}
    \put(9, -3){\textbf{GT}}
    \put(33, -3){\textbf{QEM}}
    \put(56, -3){\textbf{LpCVT}}
    \put(83, -3){\textbf{Ours}}
    \end{overpic}
    \vspace{-1pt}
    \caption{By setting the target number of vertices to 400, we compare the ability to consolidate weak features among QEM, LpCVT, and our method.
    It's noted that the original Mobius-ring model has a smoothly transitioning shape with 7000 vertices.}
    \label{fig:QEMLp}
    \vspace{-4mm}
\end{figure}

Second, while strong features are marked by drastic changes in normals within a localized area, weak features involve subtle shape variations over larger scales. For instance, in the Mobius-ring model visualized in Fig.~\ref{fig:QEMLp}, the original model exhibits a smoothly transitioning shape with dense triangulation. Our tests indicate that traditional methods like QEM and LpCVT struggle to capture these weak feature lines, unlike our algorithm, which demonstrates a superior capability in this regard.

\section{Algorithm}
Minimizing Eq.~(\ref{eq:ourObj}) resembles solving the CVT problem, which alternately performs surface decomposition and point re-location. The algorithm paradigm decouples point placement from their connections. Similar to CVT, we also utilize the L-BFGS solver to minimize Eq.~(\ref{eq:ourObj}) until some convergence criterion is satisfied.
In the following, we elaborate on (1) the computation of the objective function and gradients, (2) the utilization of decaying weight, and (3) a more effective technique for computing RVDs on thin-plate models.

\subsection{Optimization}
As mentioned above, the entire energy includes two parts denoted by $E_\text{NA}$ and $E_\text{CVT}$, respectively. The computation of the objective function and gradients relies on surface decomposition and numerical quadrature.

\paragraph{Surface Decomposition and Quadrature}
Given a set of points $\{\site_i\}_{i=1}^N$, the RVD typically decomposes the surface into $N$ regions, with each point $\site_i$ being associated with one region. However, the RVD may encounter challenges with thin-plate models, as discussed in Section~\ref{subsec:thin-plate}.
Recall that the input is represented by a triangular surface $S$. Each region, $\Omega_i$, consists of a collection of planar convex polygons. These polygons may include triangles from $S$ or sub-triangles that are sliced by the walls of the Voronoi diagram of the points $\{\site_i\}_{i=1}^N$.
To compute the contribution of a polygon with $k$ vertices, the polygon is partitioned into $k-2$ triangles. The Albrecht-Collatz quadrature is then employed to evaluate the integral over each of these triangles. Notably, the Albrecht-Collatz quadrature utilizes six points per triangle: three points are positioned at the midpoints of the edges, and the remaining three points are situated inside the triangle.

\paragraph{Gradients}
Like that achieved in CVT~\cite{liu2009centroidal},
\begin{equation}
    \nabla_{\site_i} E_\text{CVT} = \int_{\Omega_i} 
    -2M_{\text{CVT}}(\site-\site_i)
    \text{d}s.
\label{eq:GCVT}
\end{equation}
\begin{wrapfigure}{r}{2cm}
\vspace{-3.5mm}
  \hspace*{-4mm}
  \centerline{
  \includegraphics[width=25mm]{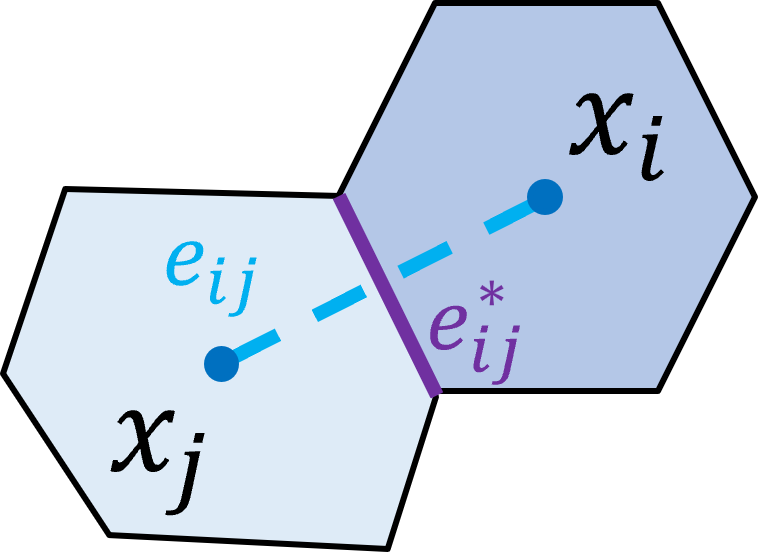}
  }
  \vspace*{-4mm}
\end{wrapfigure}
However, $\nabla_{\site_i} E_{\text{NA}}$
is not as straightforward as $\nabla_{\site_i} E_\text{CVT}$. According to Reynolds transport theorem,
we have
\begin{equation}
\begin{aligned} 
 \nabla_{\site_i} E_\text{NA} &= \int_{\Omega_i} 
    -2((\site-\site_i)\cdot  \normal_\site)\normal_\site
    \text{d}s 
\\ &+ \sum_{j\in\text{Nei}(i)}\int_{e_{ij}^*}((\site - \site_i)\cdot \normal_\site)^2(\nabla_{\site_i}\site\cdot\frac{\rve_{ij}}{\left \| \rve_{ij} \right \|}) \text{d} \site 
\\ &- \sum_{j\in\text{Nei}(i)}\int_{e_{ij}^*}((\site - \site_j)\cdot \normal_\site)^2(\nabla_{\site_i}\site\cdot\frac{\rve_{ij}}{\left \| \rve_{ij} \right \|}) \text{d} \site,
\end{aligned}
\label{eq:GNNCVT}
\end{equation}
where $\{\site_j\}_{j\in\text{Nei}(i)}$ is the point set neighboring to~$\site_i$,
$\rve_{ij}=\site_j-\site_i$,
and $\rve_{ij}^*$ is the common boundary between $\site_i$'s cell and $\site_j$'s cell (see the purple line in the wrapped figure).
According to~\cite{deGoes:2012:BNOT}, we further have:
\begin{equation}
    \nabla_{\site_i}\site \cdot\frac{\rve_{ij}}{\left \| \rve_{ij} \right \|} = \frac{\rve_{ij}}{2\| \rve_{ij}\|}.
\end{equation}

However, in our experiments, we observe that the sum of the second and third terms in Eq.~(\ref{eq:GNNCVT}) is negligible compared to the first term. Therefore, we only consider the first term to approximate the gradients:
\begin{equation}
\begin{aligned} 
 \nabla_{\site_i} E_\text{NA} &\approx \int_{\Omega_i} 
    -2((\site-\site_i)\cdot  \normal_\site)\normal_\site
    \text{d}s.
\end{aligned}
\label{eq:GNNapproximate}
\end{equation}

\begin{figure}[!tp]
    \centering
    \begin{overpic}[width=0.89\linewidth]{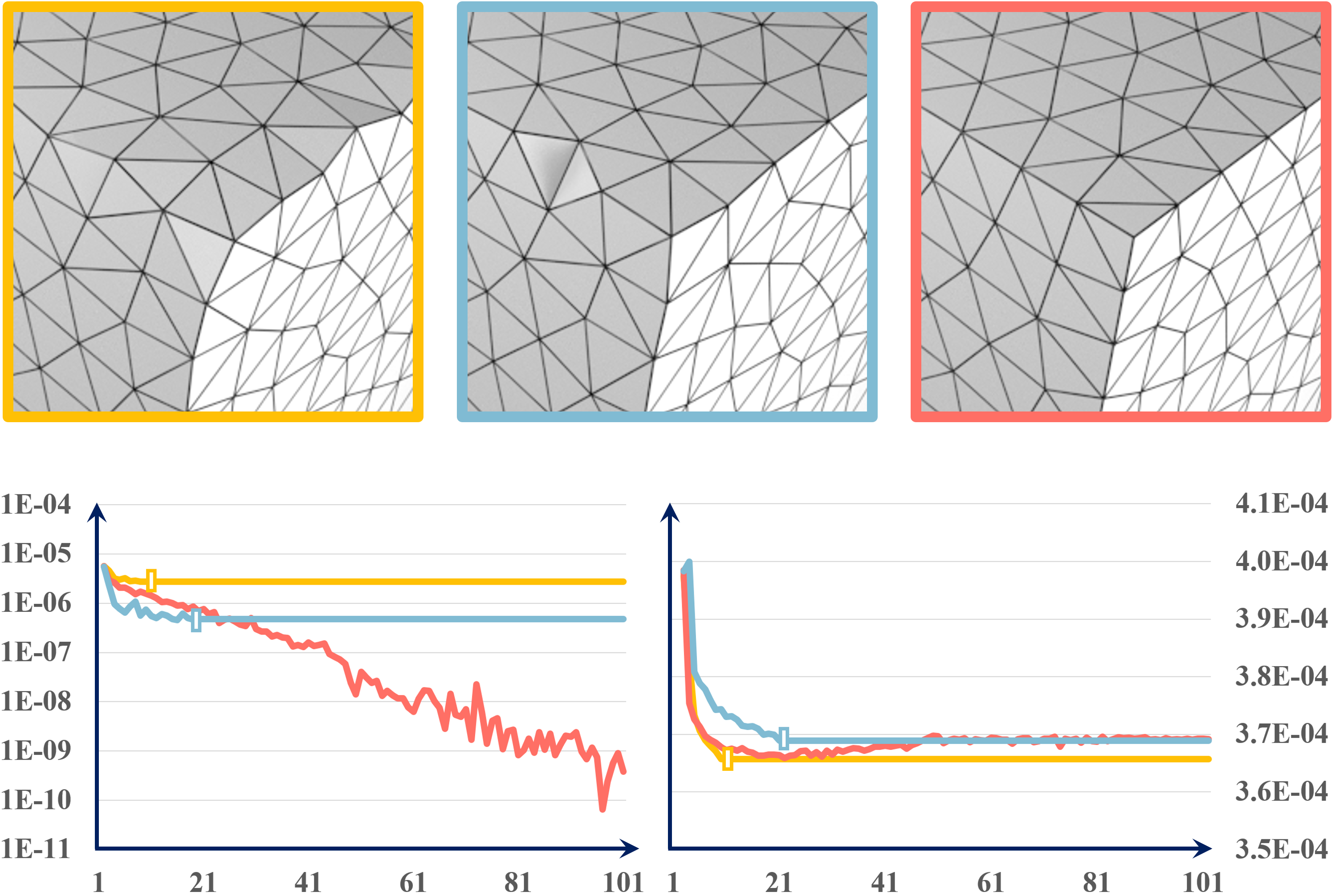}
    \put(-1.5, 32){(a) Ours w/o Decay}
    \put(40, 32){(b) LpCVT}
    \put(77, 32){(c) Ours }
    \put(12, -3.2){(d) NA Energy}
    \put(58, -3.2){(e) CVT Energy}
    \end{overpic}
    \vspace{0pt}
    \caption{
    Resulting triangle mesh surfaces.
For all three approaches, we maintain a constant total of 100 iterations and monitor changes in the normal anisotropy term (d) and the CVT energy term (e). The plots for these approaches are colored yellow, blue, and red, respectively.
Please note that the vertical axis of the normal anisotropy term is displayed on a log10 scale, whereas the vertical axis for the CVT energy term uses a linear scale. 
    }
\label{fig:FunctionValueDecay}
\vspace{-4mm}
\end{figure}
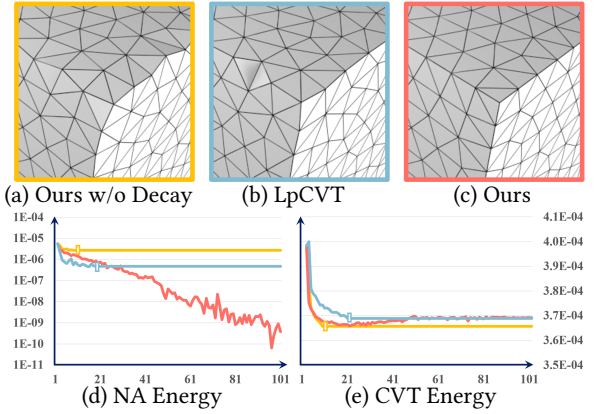

\subsection{Decaying Weight}
Suppose that the total surface area is normalized to 1.0. $E_\text{CVT}$ does not vary significantly for either CAD models or organic models. However, $E_\text{NA}$ can be reduced to nearly 0 for most CAD models, making the values of $E_\text{NA}$ and $E_\text{CVT}$ not of the same order of magnitude (to be more precise, the changes in the two terms occur at the same rate in a balanced state). For example, with the Cube model (see Fig.~\ref{fig:RVD}), $E_\text{NA}$ can be reduced to 0, which makes $E_\text{CVT}$ the dominant term. In this situation, $E_\text{CVT}$ prevents $E_\text{NA}$ from vanishing. Therefore, we need to prioritize minimizing $E_\text{CVT}$ at the start of optimization and gradually reduce its influence. To achieve this, we keep $\lambda_\text{NA}$ unchanged throughout the optimization while allowing $\lambda_\text{CVT}$ to decay according to:
\begin{equation}
    \lambda_\text{CVT}^{(i)} = \lambda_\text{CVT}^{(i-1)}  \times \decay,
\end{equation}
where $\decay=0.95$ by default. This approach prioritizes even point placement in the initial iterations and shifts focus to the feature alignment requirement subsequently. Refer to Fig.~\ref{fig:FunctionValueDecay} for plots showing the changes in these terms with and without weight decay.

The optimization generally requires tens of iterations for CAD models based on our tests. Even with up to 100 iterations, the resulting triangulation on a CAD model can still satisfy high triangle quality and feature alignment simultaneously. However, we observe that for an organic model, the triangle quality may diminish significantly as the number of iterations increases. To prevent the deterioration of triangle quality, we need to stop the optimization if the CVT energy term $E_\text{CVT}$ experiences an increase, described by:
\begin{equation}
    E_\text{CVT}^{(i)}\geq\mu \min_{j=0}^{i-1} E_\text{CVT}^{(j)},
\end{equation}
where $\mu$ is set at 1.05.

To summarize, our optimization terminates if one of the following conditions is met:
\begin{itemize}
\item The gradient norm is less than $10^{-8}$.
\item $E_\text{CVT}^{(i)} \geq 1.05 \times \min_{j=0}^{i-1} E_\text{CVT}^{(j)}.$
\end{itemize}
Using the model in Fig.~\ref{fig:Pipeline} as an example, we set the target number of movable points to 3000 and execute our algorithm. The optimization process involves 50 iterations, as the second termination condition is triggered.

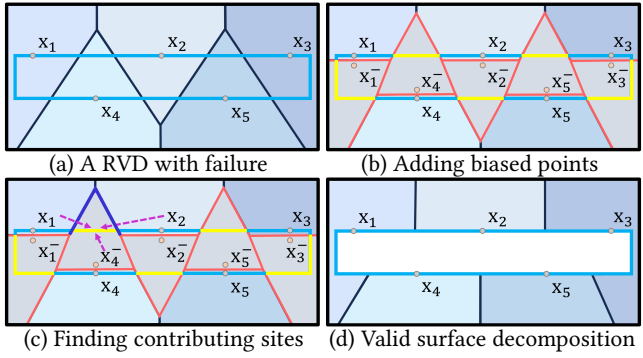
\begin{figure}[!tp]
    \centering
    \begin{overpic}[width=0.99\linewidth]{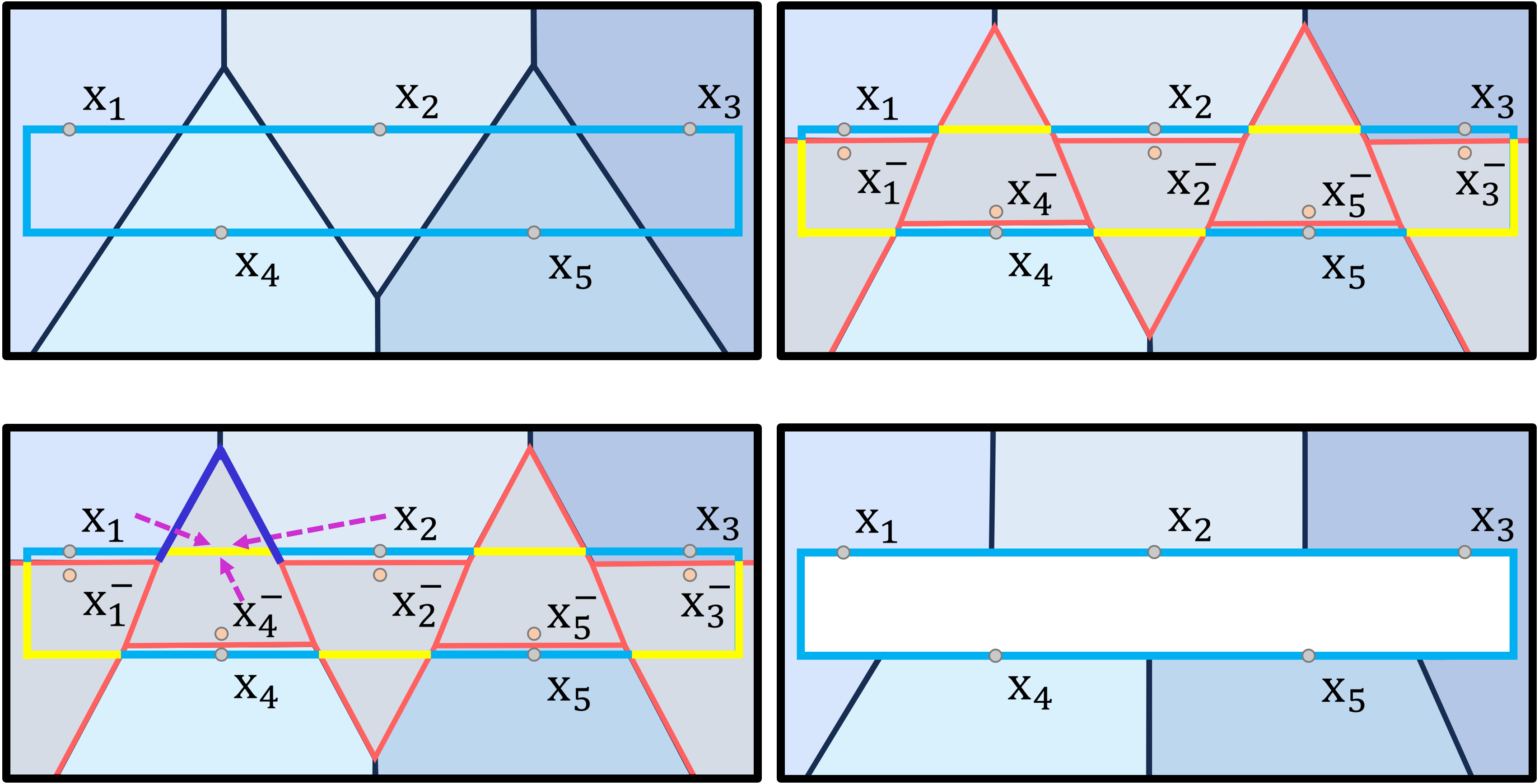}
    \put(7, 24.5){(a) A RVD with failure}
    \put(56, 24.5){(b) Adding biased points}
    \put(3, -3){(c) Finding contributing sites}
    \put(51, -3){(d) Valid surface decomposition}
    \end{overpic}
    \vspace{0pt}
    \caption{A simple yet effective technique for fixing RVDs by introducing an additional set of biased points. (a) RVDs may fail on a thin-plate model with a set of sparse sites. 
    (b) Rather than fix the problematic RVD, we bias each point at a negligible distance along the inward normal direction. 
    (c) We need to find the contributing sites for each yellow region that is dominated by a biased site.
    (d) After the yellow regions are partitioned and assigned to the contributing sites, we get a valid surface decomposition. }
    \label{fig:Thin-plate}
    \vspace{-2mm}
\end{figure}


\subsection{RVDs on Thin-plate Models}\label{subsec:thin-plate}


When the input model is a thin-plate and the number of movable points is insufficient, RVDs may encounter an issue where a single site dominates two or more regions. Some research works~\cite{wang2022restricted} focus on identifying the ownerless regions and further analyzing which sites can partition these regions, which requires tedious computation.
In this paper, we present a simple yet effective technique to address this issue. 

We use a 2D example shown in Fig.~\ref{fig:Thin-plate} to illustrate our idea, where the blue rectangle is very thin compared to the number of sites. If one directly computes the RVD, $\site_2$, $\site_4$, and $\site_5$ dominate two regions, violating the rule of ``one site, one region''; See Fig.~\ref{fig:Thin-plate}(a).
Rather than correcting the problematic RVD, we bias each point at a negligible distance~(e.g., 0.01) along the inward normal direction, yielding a point set $\{\site_i^-\}_{i=1}^N$ of equal size. 
After that, it requires two stages to accomplish the surface decomposition. 

\paragraph{Stage 1.} We compute the Voronoi diagram of $\{\site_i\}_{i=1}^N\cup \{\site_i^-\}_{i=1}^N$. The entire surface is thus decomposed into \XR{$N$ or more regions.}
The regions dominated by $\site_i^-,i=1,2,\cdots,N,$ 
 colored in yellow in Fig.~\ref{fig:Thin-plate}(b), need to be partitioned and re-assigned. 

\paragraph{Stage 2.} 
For each yellow region, we identify the ``contributing'' sites. Taking the bottom-left yellow region as an example, it is determined by two bisectors: one given by $\site_1$ and $\site_1^-$, and the other given by $\site_1^-$ and $\site_4$. We define the contributing sites of this region as $\site_1$, $\site_1^-$, and $\site_4$. By disregarding the biased site $\site_1^-$, we partition this region using $\site_1$ and $\site_4$. (Note that $\site_4$ does not gain more dominating area in this operation.) Similarly, for the top-left yellow region, the contributing sites are $\site_1$, $\site_2$, and $\site_4^-$. By disregarding the biased site $\site_4^-$, this region can be re-partitioned and assigned to $\site_1$ and $\site_2$.
See Fig.~\ref{fig:Thin-plate}(c,d).


It's worth noting that the regions dominated by biased sites are distinct from the ownerless regions in~\cite{wang2022restricted}. For instance, the bottom-left yellow region in Fig.~\ref{fig:Thin-plate}~(b) is not an ownerless region but necessitates additional handling in our algorithm. After identifying the yellow regions, we can process them separately and in parallel.
See the fixed surface decomposition in Fig.~\ref{fig:Thin-plate}~(d).

\section{Experimental Results}\label{sec:Experimental}
\begin{figure}[!h]
    \centering
    \begin{overpic}[width=0.98\linewidth]{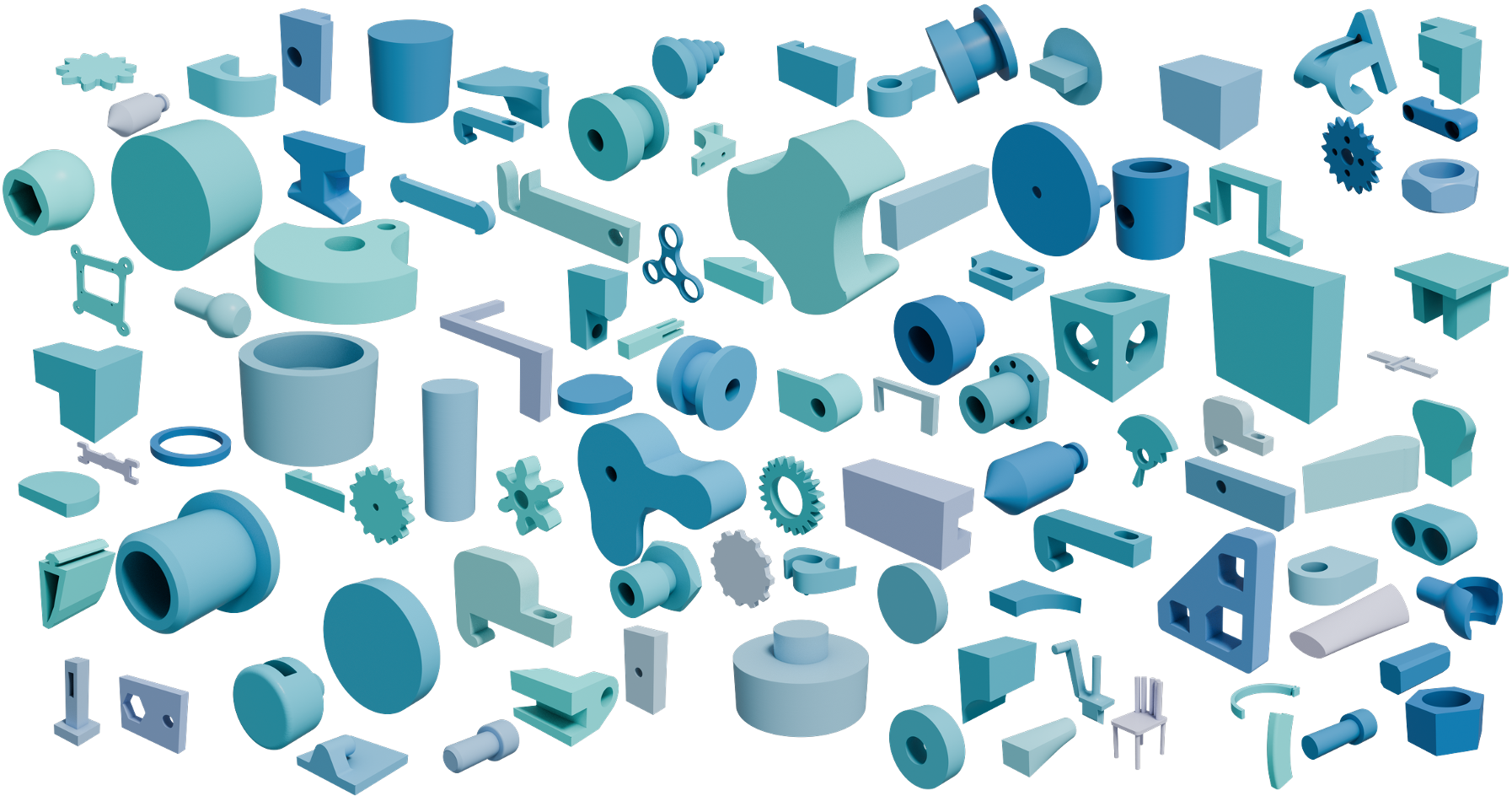}
    \end{overpic}
    \caption{
    \XR{100 models selected from the ABC dataset.}
    }
    \label{fig:dataset}
\end{figure}

All of our experiments were conducted on a computer equipped with an AMD Ryzen 9 5950X CPU and 64 GB of memory. The ABC dataset~\cite{koch2019abc} includes models with self-intersections, non-manifold vertices/edges, and open boundaries. Therefore, we randomly selected 100 watertight manifold meshes, similar to those in RFEPS~\cite{xu2022rfeps}. See Fig.~\ref{fig:dataset} for a gallery of the selected models. Additionally, we selected 21 organic models with weak features.

For each CAD model, we set the target number of vertices to be 500, 1000, and 2000, respectively. For organic models, we sampled various target numbers of vertices, ranging from 100 to 8000.

\paragraph{Evaluation Metrics}
To measure the difference between the simplified surface and the original version, we use four indicators: \textit{Chamfer Distance} (CD),  \textit{F-score} (F1), \textit{Normal Consistency} (NC), \XR{and \textit{Hausdorff Distance} (HD).  

To facilitate the definition of metrics, we define $M_g$ and $M_p$ as the ground truth model and the simplified model, respectively. Let $\mathcal{X} \subset M_g$ and $\mathcal{Y} \subset M_p$ denote the randomly sampled points from each model, respectively. We sample 100K points for all evaluations. The nearest point can be found as follows:
\begin{equation}
\begin{aligned}
\mathcal{P}_{g 2 p}(\mathbf{x})=\arg \min _{\mathbf{y} \in \mathcal{Y}}\|\mathbf{x}-\mathbf{y}\|_2, \\
\mathcal{P}_{p 2 g}(\mathbf{y})=\arg \min _{\mathbf{x} \in \mathcal{X}}\|\mathbf{x}-\mathbf{y}\|_2.
\end{aligned}
\label{eq:qp}
\end{equation}
}

%

\XR{
\textit{Chamfer Distance} (CD). To measure the average nearest squared distance between $\mathcal{Y}$ and $\mathcal{X}$, we introduce the chamfer distance metric as follows:

\begin{equation}
\begin{aligned}
\mathrm{CD} (\mathcal{X}, \mathcal{Y}) &= \frac{1}{2 N_g} \sum_{i=1}^{N_g}\|\mathbf{x}_i-\mathcal{P}_{g 2 p}(\mathbf{x}_i)\|_2+\\ 
&\frac{1}{2 N_p} \sum_{i=1}^{N_p}\|\mathbf{y}_i-\mathcal{P}_{p 2 g}(\mathbf{y}_i)\|_2.
\label{eq:cd}
\end{aligned}
\end{equation}

\textit{F-score} (F1).
F-score is the harmonic mean of precision and recall. Precision is the ratio of true positives to the sum of true positives and false positives, while recall is the ratio of true positives to the sum of true positives and false negatives.
\begin{equation}
\text{F1}(\mathcal{X}, \mathcal{Y})=\frac{2 * \text { precision } * \text { Recall }}{\text { precision }+\text { Recall }}.
\label{eq:fs}
\end{equation}

\textit{Normal Consistency} (NC).
To measure the alignment of normals, we define a normal consistency score by calculating the average of the absolute dot products between pairs of normals.
\begin{equation}
\begin{aligned}
\mathrm{NC}(\mathcal{X}, \mathcal{Y})=& \frac{1}{2 N_g} \sum_{i=1}^{N_g}|\mathbf{n}(\mathbf{x}_i) \cdot \mathbf{n}(\mathcal{P}_{g 2 p}(\mathbf{x}_i))|+ \\
&\frac {1}{2 N_p} \sum_{i=1}^{N_p}|\mathbf{n}(\mathbf{y}_i) \cdot \mathbf{n}(\mathcal{P}_{p 2 g}(\mathbf{y}_i))|
\end{aligned}
\label{eq:nc}
\end{equation}

\textit{Hausdorff Distance} (HD). The Hausdorff distance is a distance metric that can measure the distance between two subsets within the same metric space. $\mathop{HD} \limits ^{\longrightarrow}\left(\mathcal{X}, \mathcal{Y}\right)$ defines the one-sided Hausdorff distance from $\mathcal{X}$ to $\mathcal{Y}$, and $\mathrm{HD}\left(\mathcal{X}, \mathcal{Y}\right)$ is the two-sided version. 
\begin{equation}
\begin{aligned}
&\mathop{HD} \limits ^{\longrightarrow}\left(\mathcal{X}, \mathcal{Y}\right)=\max_{\mathbf{x} \in \mathcal{X}} \min _{\mathbf{y} \in \mathcal{Y}}\left\|\mathbf{x}-\mathbf{y}\right\|_2,\\
&\mathrm{HD}\left(\mathcal{X}, \mathcal{Y}\right)=\max \left(\mathop{HD} \limits ^{\longrightarrow}\left(\mathcal{X}, \mathcal{Y}\right),
\mathop{HD} \limits ^{\longrightarrow}\left(\mathcal{Y}, \mathcal{X}\right)\right).
\end{aligned}
\end{equation}
}

To evaluate triangle quality, we use the \textit{TriangleQ} indicator~\cite{frey1999surface}. A value closer to 1.0 indicates that the triangle is nearly equilateral.
\begin{equation}
\text{\textit{TriangleQ}}(t) = \frac{6}{\sqrt{3}}\frac{S_t}{p_t h_t}
\end{equation}
where $S_t$, $p_t$, and $h_t$ represent the area, half-perimeter, and the longest edge length of the triangle $t$, respectively. Furthermore, we use \textit{OpenB} to represent the number of open mesh edges in the simplification result, and \textit{NMV} to denote the number of non-manifold vertices.
For CAD models, we also employ \textit{Edge Chamfer Distance} (ECD) and \textit{Edge F-score} (EF1), proposed by NMC~\cite{chen2021nmc}, to measure the extent to which the feature lines are preserved.


\begin{table}[!h]
\caption{Quantitative comparison on $100$ CAD models, taken from the ABC dataset~\cite{koch2019abc}.
Each CAD model has strong features.
The \underline{\textbf{best}} scores are emphasized in bold with underlining, while the \textbf{second best} scores are highlighted only in bold.
}
\vspace{-2mm}
\label{tab:comp_cad}
\resizebox{\linewidth}{!}{
\begin{tabular}{c|c|ccccccccccc}
\toprule
\multicolumn{1}{l|}{}  & $\#V$             & CVT   & LpCVT & QEM   & SMS   & IEM   & LPM   & PQT   & PQP   & MD   & ERB   & Ours    \\ \midrule
\multirow{3}{*}{$\mathrm{CD}\left(\times 10^{4}\right) \downarrow$}  
                       & 500           & 0.349 & 0.148 & 0.119 & 0.335 & 2.876 & 0.126 & 0.573 & \textbf{0.099} & 16.650 & \under{\textbf{0.098}} & 0.115 \\
                       & 1000          & 0.174 & 0.098 & 0.108 & 0.146 & 0.881 & 0.127 & 0.262 & \under{\textbf{0.085}} & 16.339 & 0.098 & \textbf{0.093} \\
                       & 2000          & 0.110 & 0.080 & 0.090 & 0.123 & 0.267 & 0.116 & 0.155 & \textbf{0.078} & 13.816 & 0.098 & \under{\textbf{0.076}}\\ \cmidrule{1-13}
\multirow{3}{*}{$\mathrm{F1}\uparrow$}    
                       & 500           & 0.743 & 0.893 & \under{\textbf{0.933 }}& 0.804 & 0.784 & 0.908 & 0.605 & 0.908 & 0.675 & 0.920 & \textbf{0.929} \\
                       & 1000          & 0.823 & 0.923 & \under{\textbf{0.940}} & 0.878 & 0.863 & 0.907 & 0.750 & 0.930 & 0.695 & 0.920 & \textbf{0.938} \\
                       & 2000          & 0.884 & 0.938 & \textbf{0.943} & 0.901 & 0.914 & 0.923 & 0.842 & \textbf{0.943} & 0.738 & 0.920 & \under{\textbf{0.944}} \\ \cmidrule{1-13}
\multirow{3}{*}{$\mathrm{NC}\uparrow$}    
                       & 500           & 0.947 & 0.978 & 0.985 & 0.982 & 0.928 & 0.982 & 0.941 & 0.985 & 0.841 & \textbf{0.986} & \under{\textbf{0.987}} \\
                       & 1000          & 0.964 & 0.986 & 0.986 & 0.987 & 0.967 & \textbf{0.988} & 0.958 & \textbf{0.988} & 0.847 & 0.986 & \under{\textbf{0.989}} \\
                       & 2000          & 0.975 & \textbf{0.989} & 0.988 & \textbf{0.989} & 0.982 & \textbf{0.989} & 0.969 & \textbf{0.989} & 0.867 & 0.986 & \under{\textbf{0.990}} \\ \cmidrule{1-13}
\multirow{3}{*}{$\mathrm{ECD}\left(\times 10^{2}\right)\downarrow$}   
                       & 500           & 4.247 & 0.120 & \textbf{0.105} & 3.716 & 1.577 & 0.839 & 11.340 & 2.762 & 10.250 & 0.137 & \under{\textbf{0.101}}\\
                       & 1000          & 2.600 & \textbf{0.080} & 0.088 & 0.539 & 1.360 & 0.431 & 11.598 & 0.198 & 10.664 & 0.137 & \under{\textbf{0.079}} \\
                       & 2000          & 2.560 & 0.061 & \under{\textbf{0.045}} & 0.152 & 0.765 & 0.448 & 11.115 & 0.135 &  9.001 & 0.137 & \textbf{0.055} \\ \cmidrule{1-13}
\multirow{3}{*}{$\mathrm{EF1}\uparrow$}   
                       & 500           & 0.020 & 0.461 & \under{\textbf{0.573}} & 0.208 & 0.383 & 0.392 & 0.004 & 0.563 & 0.161 & 0.560 & \textbf{0.566}\\
                       & 1000          & 0.056 & 0.553 & \under{\textbf{0.607}} & 0.387 & 0.444 & 0.468 & 0.006 & 0.586 & 0.186 & 0.560 & \textbf{0.601} \\
                       & 2000          & 0.108 & 0.593 & \under{\textbf{0.622}} & 0.487 & 0.510 & 0.520 & 0.011 & 0.596 & 0.230 & 0.560 & \textbf{0.615}\\ \cmidrule{1-13}
\multirow{3}{*}{$TriangleQ \uparrow$}   
                       & 500           & \under{\textbf{0.879}} & 0.840 & 0.525 & 0.767 & 0.483 & 0.261 & 0.761 & 0.579 & 0.407 & 0.766 & \textbf{0.857}\\
                       & 1000          & \under{\textbf{0.892}} & 0.859 & 0.573 & 0.766 & 0.538 & 0.274 & 0.754 & 0.602 & 0.468 & 0.766  & \textbf{0.883}\\
                       & 2000          & \textbf{0.892} & 0.872 & 0.641 & 0.769 & 0.626 & 0.305 & 0.740 & 0.632 & 0.534 & 0.766 & \under{\textbf{0.893}} \\ \cmidrule{1-13}
\multirow{3}{*}{$\mathrm{HD}\left(\times 10^{2}\right)\downarrow$} 
                       & 500           & 5.542 & 4.461 & 0.243 & 2.064 & 3.279 & 1.103 & 0.543 & \under{\textbf{0.036}} & 2.261 & \textbf{0.083} & 0.201 \\
                       & 1000          & 2.487 & 3.706 & \textbf{0.042} & 0.838 & 1.397 & 0.952 & 0.312 & \under{\textbf{0.021}} & 2.160 & 0.083 & 0.062 \\
                       & 2000          & 1.296 & 1.923 & 0.034 & 0.442 & 0.436 & 0.790 & 0.184 & \under{\textbf{0.011}} & 1.811 & 0.083  & \textbf{0.020}\\ \cmidrule{1-13}
\multirow{3}{*}{$OpenB \downarrow$} & 500           & \under{\textbf{0}} & \under{\textbf{0}} & 346 & 247 & 285 & 24 & \textbf{4} & 5 & 683 & 166 & \under{\textbf{0}} \\
                       & 1000          & \under{\textbf{0}} & \under{\textbf{0}} & 509 & 384 & 467 & 43 & \textbf{2} & 9 & 1213 & 166 & \under{\textbf{0}}  \\
                       & 2000          & \under{\textbf{0}} & \under{\textbf{0}} & 660 & 576 & 638 & 38 & \textbf{2} & 10 & 2198 & 166 & \under{\textbf{0}}  \\ \cmidrule{1-13}

\multirow{3}{*}{$NMV \downarrow$}   & 500           & \under{\textbf{0}} & \under{\textbf{0}} & 312 & 69 & 720 & \under{\textbf{0}} & 2 & \textbf{1} & 238 & 243 & \under{\textbf{0}}  \\
                       & 1000          & \under{\textbf{0}} & \under{\textbf{0}} & 661 & 242 & 1348 & \under{\textbf{0}} & 2 & \textbf{1} & 321 & 243 & \under{\textbf{0}}  \\
                       & 2000          & \under{\textbf{0}} & \under{\textbf{0}} & 1035 & 696 & 2342 & \under{\textbf{0}} & \textbf{2} & \textbf{2} & 440 & 243 & \under{\textbf{0}}  \\ \bottomrule
\end{tabular}
}
\vspace{-4mm}
\end{table}

\begin{table}[!h]
\caption{Quantitative comparison on $21$ organic models with weak features.}
\vspace{-3mm}
\label{tab:comp_organic}
\resizebox{\linewidth}{!}{
\begin{tabular}{c|ccccccccccc}
\toprule
\multicolumn{1}{l|}{} & CVT   & LpCVT & QEM   & SMS   & IEM   & LPM   & PQT   & PQP   & MD   & ERB   & Ours    \\ \midrule
{$\mathrm{CD}\left(\times 10^{4}\right) \downarrow$}  
                      & 0.201 & 0.106 & \under{\textbf{0.063}} & 0.301 & 0.268 & 0.093 & 0.332 & 0.105 & 0.198 & \textbf{0.069} & 0.090\\  
{$\mathrm{F1}\uparrow$}
                      & 0.776 & 0.876 & \under{\textbf{0.948}} & 0.748 & 0.850 & 0.894 & 0.754 & 0.934 & 0.901  & \textbf{0.942} &0.902\\  
{$\mathrm{HD}\left(\times 10^{2}\right) \downarrow$}  
                      & 0.829 & 0.575 & 0.110 & 0.259 & 0.710 & 0.540 & 0.344 & \under{\textbf{0.098}} & 0.170 & \textbf{0.103} & 0.147 \\ 
{$\mathrm{NC}\uparrow$}
                      & 0.962 & 0.973 & 0.935 & 0.822 & 0.973 & \under{\textbf{0.976}}  & 0.966 & 0.960 & 0.962 & 0.972 & \under{\textbf{0.976}} \\  \midrule
$TriangleQ \uparrow$                   & \under{\textbf{0.892}} & 0.847 & 0.531 & 0.611 & 0.684 & 0.478 & 0.674  & 0.518 & 0.486 & 0.781 & \textbf{0.861} \\  
$OpenB \downarrow$                & \under{\textbf{0}} & \under{\textbf{0}} & 18 & 1 & 478 & 3 & 2 & 69 & 38 & 0 & \under{\textbf{0}} \\  
$NMV \downarrow$                  & \under{\textbf{0}} & \under{\textbf{0}} & 2 & 3 & 3 & \under{\textbf{0}} & 4 & 4 & 2 & 1 & \under{\textbf{0}} \\ \bottomrule
\end{tabular}
}
\vspace{-4mm}
\end{table}

\paragraph{Parameters}
In our experiments, we use the same parameter settings for all the models used in this paper, whether CAD models or organic models. We set $\lambda_\text{NA} = \lambda_\text{CVT} = 1.0$ at the beginning of our optimization. $\lambda_\text{NA}$ remains unchanged throughout the optimization, but $\lambda_\text{CVT}$ undergoes a decaying process at the rate of $\tau = 0.95$. The parameter $\mu$, which controls the extent of the rise in the CVT energy, is set to $1.05$. We employ the L-BFGS solver to solve the optimization, and the termination condition is set by referring to the gradient norm, with a tolerance of $1e^{-8}$. 
We show an ablation study of these parameters in Section~\ref{sec:ablation}.


\begin{figure*}[!t]
    \centering
    \begin{overpic}[width=\linewidth]{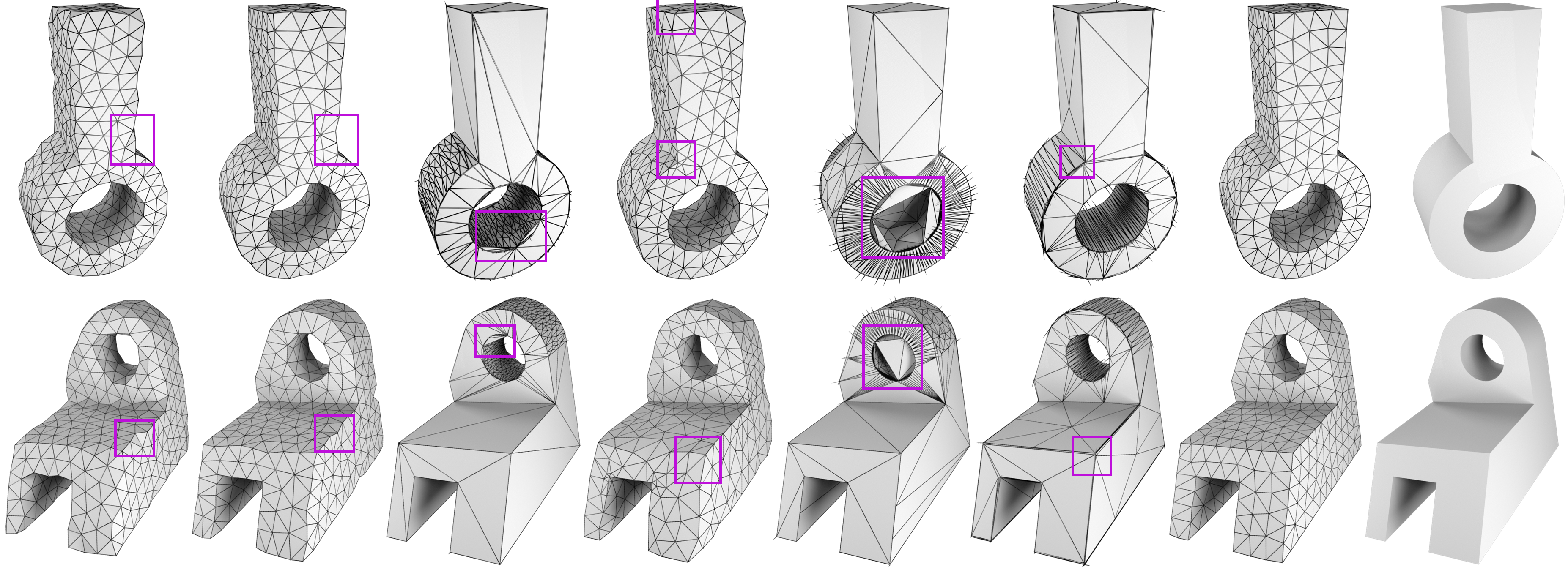}
\put( 4,-1.5){\textbf{CVT}}
\put(16,-1.5){\textbf{LpCVT}}
\put(30,-1.5){\textbf{QEM}}
\put(43,-1.5){\textbf{SMS$^\ast$}}
\put(55,-1.5){\textbf{IEM}}
\put(67,-1.5){\textbf{LPM}}
\put(79,-1.5){\textbf{Ours}}
\put(92,-1.5){\textbf{GT}}
    \end{overpic}
    \caption{
    Comparison with state-of-the-art methods on two CAD inputs. Our method excels in both accuracy and manifoldness. The target number of vertices for both inputs is set at 500.
    Note that SMS deteriorates when the input surface contains open boundaries, and we show their results on manifold inputs.
    }
    \label{fig:comp_CAD}
\end{figure*}

\begin{figure}[!htp]
    \centering
    \begin{overpic}[width=\linewidth]{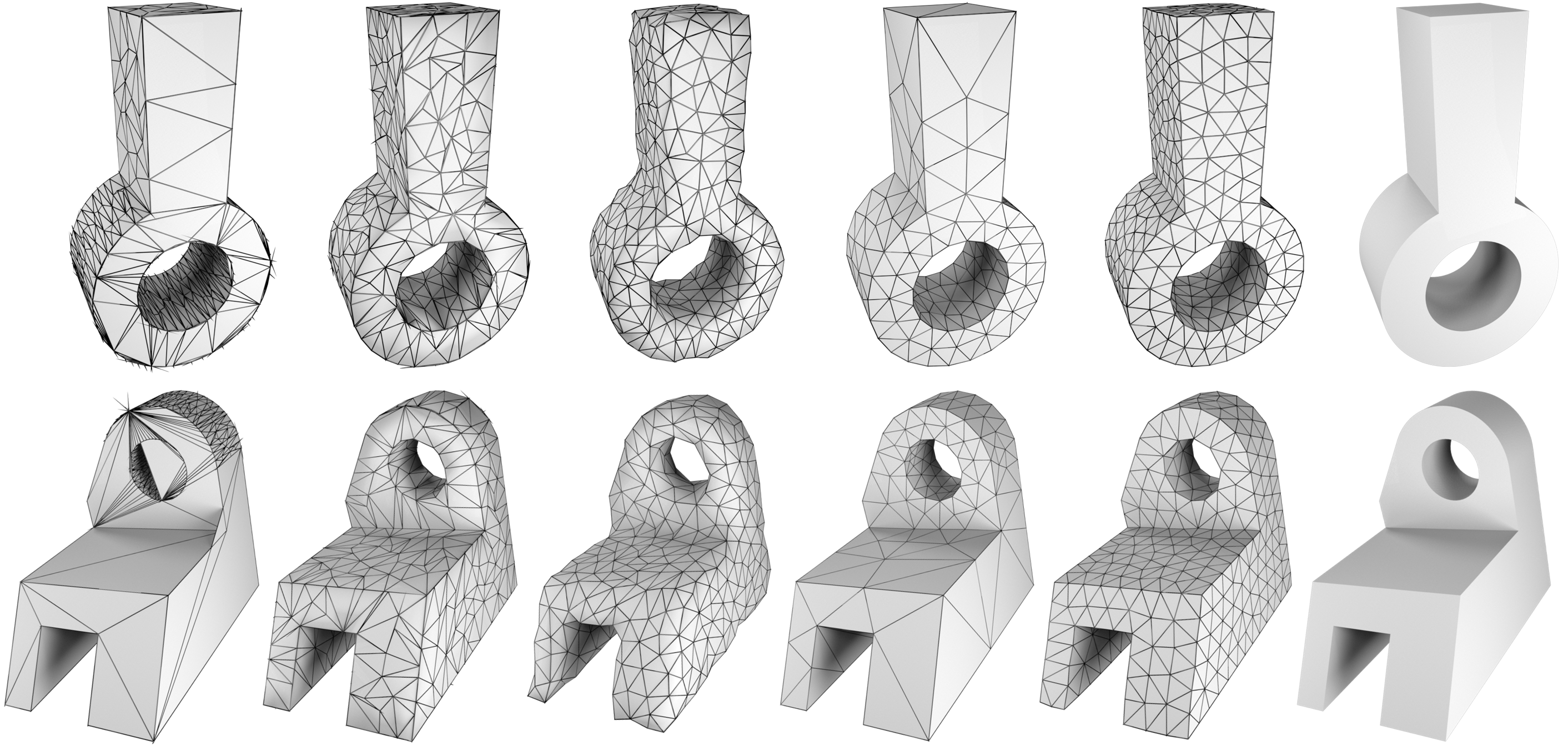}
\put( 4,-2.5){\textbf{MD$^\ast$}}
\put(20,-2.5){\textbf{PQP}}
\put(37,-2.5){\textbf{PQT}}
\put(55,-2.5){\textbf{ERB$^\ast$}}
\put(70,-2.5){\textbf{Ours}}
\put(88,-2.5){\textbf{GT}}
    \end{overpic}
    \caption{
    Comparison with state-of-the-art methods on two CAD inputs. The target number of vertices for both inputs is set at 500.
    Note that MD and ERB deteriorate when the input surface contains open boundaries, and we show their results on manifold inputs.
    }
    \vspace{-4mm}
    \label{fig:comp_CAD2}
\end{figure}

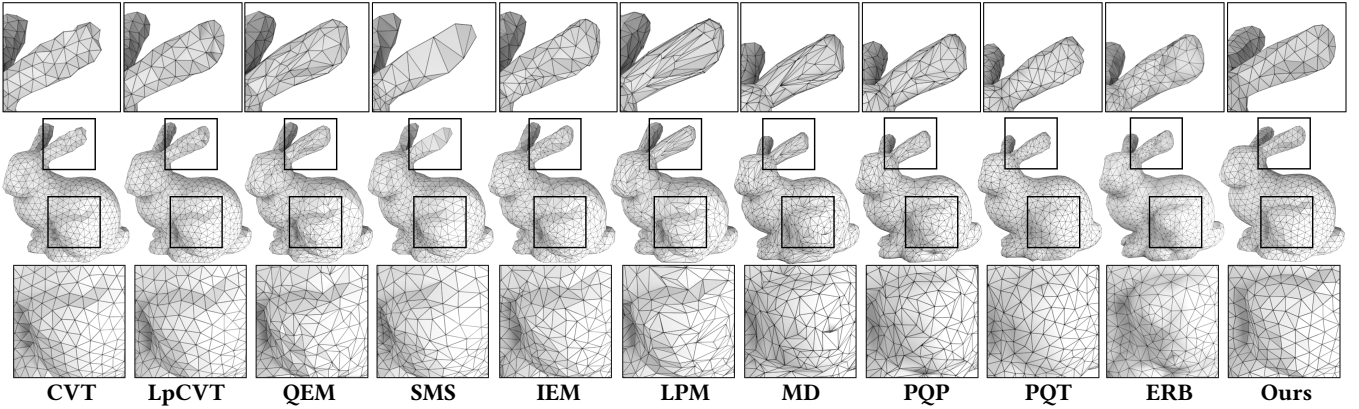
\begin{figure*}[!t]
    \centering
    \begin{overpic}[width=\linewidth]{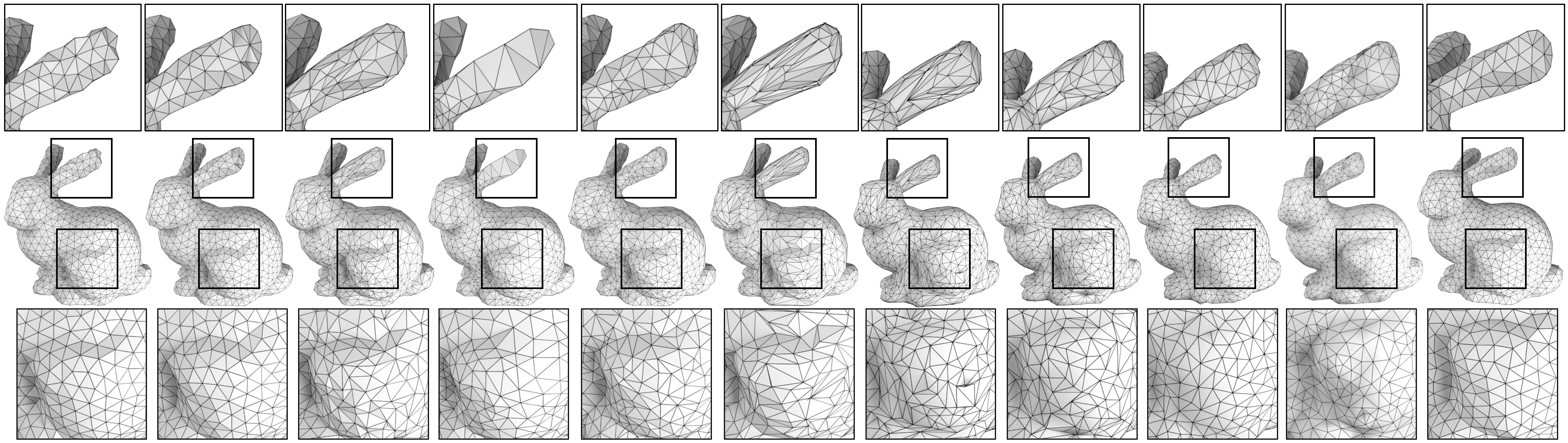}
\put(3.5,-1.5){\textbf{CVT}}
\put(11,-1.5){\textbf{LpCVT}}
\put(21,-1.5){\textbf{QEM}}
\put(30.5,-1.5){\textbf{SMS}}
\put(39.8,-1.5){\textbf{IEM}}
\put(49,-1.5){\textbf{LPM}}
\put(58,-1.5){\textbf{MD}}
\put(67,-1.5){\textbf{PQP}}
\put(76,-1.5){\textbf{PQT}}
\put(85,-1.5){\textbf{ERB}}
\put(93.5,-1.5){\textbf{Ours}}

    \end{overpic}
    \caption{
    Comparison with state-of-the-art methods on the bunny model using 1000 sample points, demonstrating that our method can more effectively consolidate weak features.    
    }
    \label{fig:comp_organic}
\end{figure*}

\subsection{Comparison Methods}
\label{sec:comparison}
We compared our method with nine state-of-the-art (SOTA) methods: CVT~\cite{du1999centroidal}, LpCVT~\cite{levy2010p}, QEM~\cite{garland1997surface}, SMS~\cite{lescoat2020spectral}, IEM~\cite{liu2023surface},  LPM~\cite{chen2023robust}, 
\XR{PQ~\cite{trettner2020fast},
MD~\cite{kobbelt1998general} and ERB~\cite{hu2016error}.}
In the following, we will briefly introduce each method and its parameter settings.

CVT~\cite{du1999centroidal} optimizes site placement from an energy perspective, achieving high-quality triangulations. LpCVT~\cite{levy2010p} extends CVT to better preserve mesh features. In the default setting, the coefficient of normal anisotropy is set to 4. 
QEM~\cite{garland1997surface} simplifies input meshes by minimizing the quadratic error metric. QEM supports the user-specified number of faces. 
SMS~\cite{lescoat2020spectral} is a spectrum-based meshing method that takes the number of preserved eigenvalues as input, which we set to 50 after experimentation.
IEM~\cite{liu2023surface} extends QEM to preserve input model features from an intrinsic perspective.
LPM~\cite{chen2023robust} extracts a low-resolution isosurface with features and progressively guides the simplification of the original surface using QEM to align with the input. LPM requires the resolution parameter to extract the isosurface, set to 100 in our experiments.
\XR{
PQ~\cite{trettner2020fast} revolutionizes error quadric minimization by embedding it in a probabilistic framework, enabling robust and efficient solutions that are up to 50 times faster than traditional SVD methods and enhance mesh processing tasks with greater uniformity and noise resilience. In the implementation of PQ~\cite{trettner2020fast}, two distinct configurations are offered: \textit{`prob\_triangle'} and \textit{`prob\_plane'}, which we denote as PQT and PQP, respectively, to differentiate their applications in probabilistic quadric error minimization.
MD~\cite{kobbelt1998general} establishes a generalized framework for mesh reduction, similar to greedy heuristic optimization algorithms, providing a versatile template that adapts to the specific needs of various target applications.
ERB~\cite{hu2016error} introduces a novel surface remeshing algorithm that adeptly balances geometric fidelity, minimal complexity, and quality by simultaneously optimizing for an exact approximation error bound, minimal interior angles, and vertex count, resulting in superior meshes for geometry processing applications. It is important to note that ERB~\cite{hu2016error} does not support the specification of target points, which sets it apart from other methods.
}

\begin{figure*}[!htp]
    \centering
    \begin{overpic}[width=\linewidth]{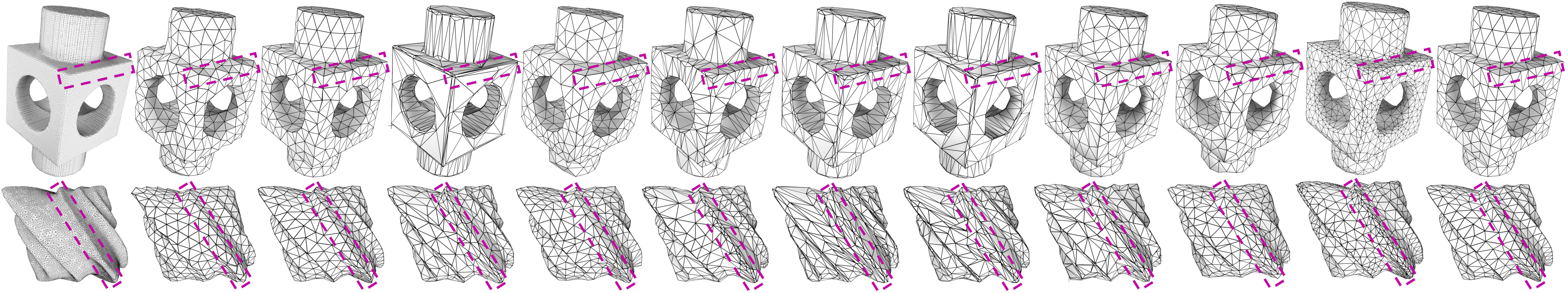}
\put( 0,-1.5){\textbf{Base Surface}}
\put(12,-1.5){\textbf{CVT}}
\put(17.5,-1.5){\textbf{LpCVT}}
\put(27,-1.5){\textbf{QEM}}
\put(36,-1.5){\textbf{SMS}}
\put(44,-1.5){\textbf{IEM}}
\put(52,-1.5){\textbf{LPM}}
\put(61,-1.5){\textbf{MD}}
\put(69,-1.5){\textbf{PQP}}
\put(77,-1.5){\textbf{PQT}}
\put(85.5,-1.5){\textbf{ERB}}
\put(93.5,-1.5){\textbf{Ours}}
    \end{overpic}
    \caption{
   Comparison with state-of-the-art methods on two smoothed inputs. The target number of points is set at 500. Our method more effectively consolidates weak features, whereas the QEM family may result in a chamfer along the feature line.
    }
    \label{fig:comp_weak}
\end{figure*}
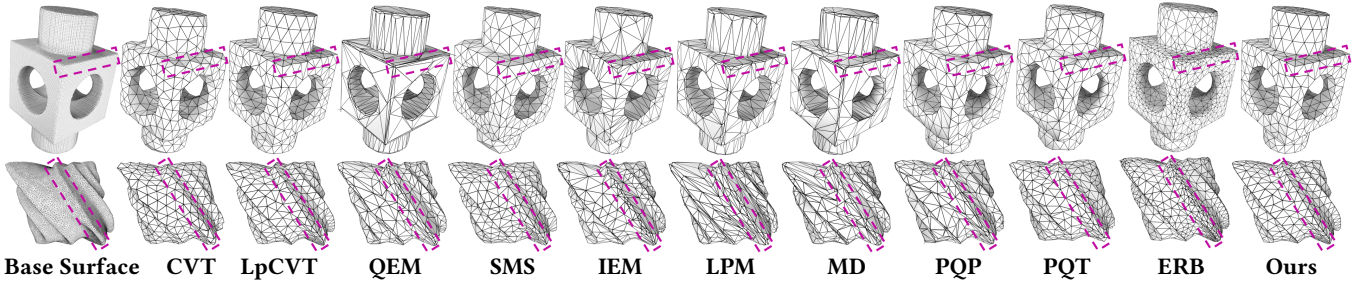


\subsection{Comparisons on CAD Models}
\label{sec:comp_cad}
We present the statistics of quantitative comparisons for the $100$ CAD models in Table~\ref{tab:comp_cad}. The qualitative comparisons are illustrated in Fig.~\ref{fig:comp_CAD} and Fig.~\ref{fig:comp_CAD2}. It is evident that the CVT method excels in triangulation quality but falls short in preserving features. The LpCVT outperforms CVT in terms of the ability to preserve sharp features. In comparison with CVT and LpCVT, our method consistently demonstrates lower CD/HD scores and higher F1/NC scores, regardless of the number of input sites ($500$, $1000$, and $2000$), which validates the advantage of our method in preserving sharp features.

As the number of points increases, QEM excels in accuracy preservation, as indicated by the ECD metric in Table~\ref{tab:comp_cad}. However, the \textit{TriangleQ} indicator reveals that QEM cannot ensure high triangulation quality, with a significant number of obtuse triangles. SMS, IEM, LPM, PQP and ERB also demonstrate an accuracy advantage in terms of the CD, F1 and NC scores but may suffer from various artifacts, such as open mesh edges and non-manifold vertices. In comparison with these methods, our approach simultaneously addresses the requirements of accuracy, triangle quality, and feature alignment.

\XR{
Note that PQT (referred to as `prob\_triangle' in PQ~\cite{trettner2020fast}) demonstrates significant enhancement in triangulation quality over PQP (referred to as `prob\_plane' in PQ~\cite{trettner2020fast}), albeit at the expense of accuracy, particularly for sharp feature indicators like ECD and EF1. ERB~\cite{hu2016error} excels across various metrics; however, its inability to specify point count often results in a higher average number of points used, approximately 874 in our analysis. Additionally, ERB is prone to issues with open boundaries and non-manifolds.
}

\subsection{Comparisons on Organic Models}
\label{sec:comp_organicc}
We also test our method on $21$ organic models with weak features and show the quantitative statistics in Table~\ref{tab:comp_organic}.
Similar to CAD model comparisons discussed in Section~\ref{sec:comp_cad},  
CVT and LpCVT achieve excellent triangulation quality but exhibit lower accuracy. 
QEM, SMS, IEM, LPM, PQP, MD and ERB are better than CVT and LpCVT in terms of simplification accuracy, by the cost of 
diminishing triangle quality. 
It can be seen from Table~\ref{tab:comp_organic} that our approach consistently achieves optimal or near-optimal scores, whether in accuracy or triangulation quality.



Due to the nature of organic models, some inherent weak features are not as distinctive as strong features. These weak features are prone to being erased in the simplification result. In Fig.~\ref{fig:comp_organic} and Fig.~\ref{fig:comp_weak}, it is evident that our method can consolidate weak features while maintaining high-quality triangulation, like the ears, legs, and neck of the \textit{bunny}. With this advantageous property, our algorithm has the potential to convey visual shape clues through a simplified representation, distinguishing itself from other mesh simplification algorithms.

\begin{figure}[!h]
\centering
\begin{overpic}
[width=0.99\linewidth]{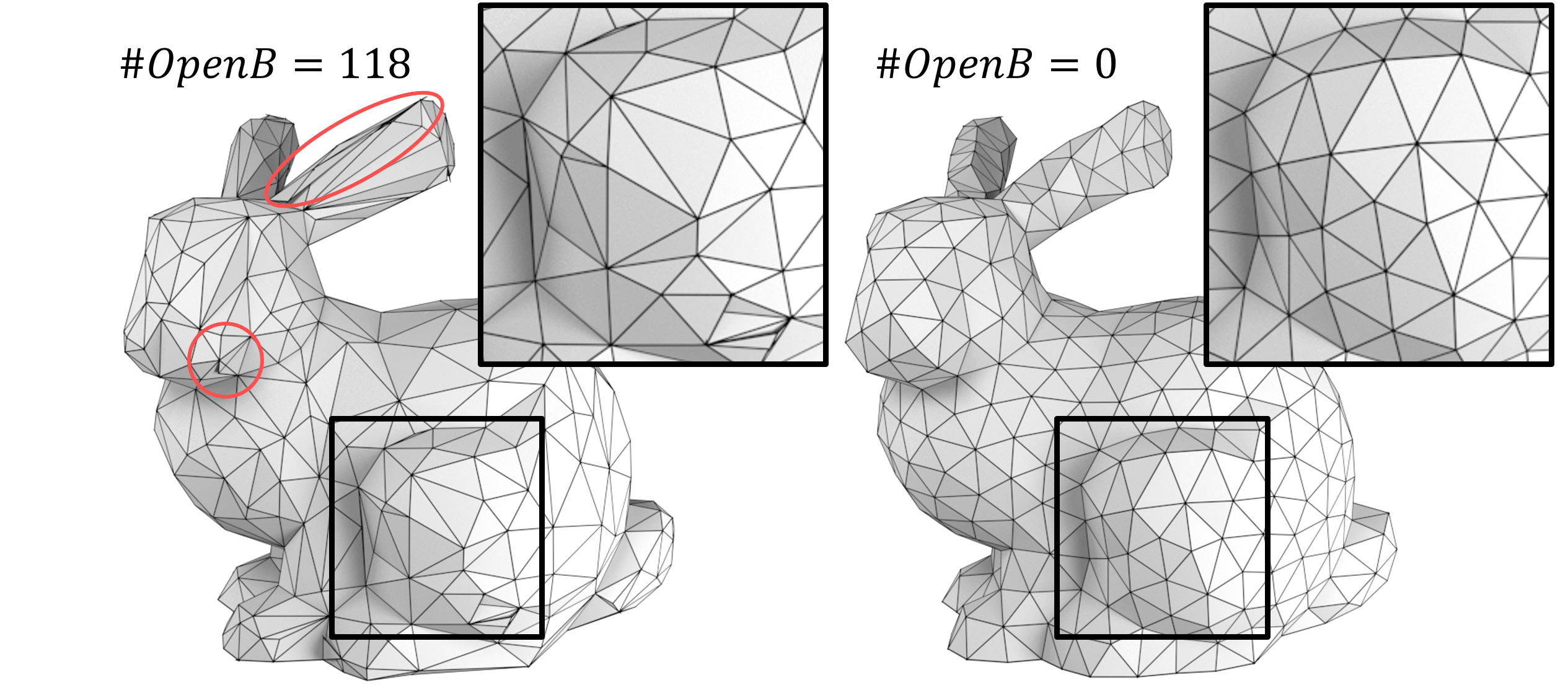}
\put(7, -3) {VSR~\cite{zhao2023variational}}
\put(75, -3){Ours}

\end{overpic}
\caption{
Comparison with VSR~\cite{zhao2023variational}. Due to VSR is a QEM-based method that involves a mixed-integer programming step and practical success in finding a valid solution is not guaranteed. Numerous open mesh edges exist in this example due to the failure to satisfy manifold-edge constraints.
}
\label{fig:VSR}
\end{figure}

Additionally, VSR~\cite{zhao2023variational} can be employed for mesh simplification by transforming the input mesh into a set of points. VSR alternates between point clustering and relocation. As the point relocation operation follows the spirit of QEM, the resulting point distribution may not be even when the number of points is limited.
In Fig.~\ref{fig:VSR}, with the number of points set to 500 (the maximum supported by VSR), VSR produces a simplified mesh with poor triangulation. 
Furthermore, VSR involves a mixed-integer programming step, and practical success in finding a valid solution is not guaranteed. Consequently, numerous open mesh edges exist in this example due to the failure to satisfy manifold-edge constraints.



\subsection{Run-time Performance}
\label{sec:runtime}
We present the run-time performance statistics in Table~\ref{tab:time}. The tests were conducted on the block model and bunny model, each with varying resolutions ranging from $0.5K$ points to $10K$ points. The total running time primarily includes the construction of RVDs and the optimization, with the computation of RVDs being the most time-consuming operation.
Additionally, it can be observed that the optimization typically requires 40 to 50 iterations. The practical timing cost is approximately $O\left(k(m+n)\right)$, where $m$ is the complexity of the base mesh, $n$ is the target number of vertices, and $k$ is the number of optimization iterations.

\begin{table*}[!h]
\caption{Running time (in seconds) with respect to the number of sampling points~(\#V). The input surface mesh of these two models contains both 40K faces. Note that CVT, LpCVT, and our method are based on RVD computation, so we show the running time as L-BFGS time / RVD time. 
}
\vspace{-3mm}
\label{tab:time}
\resizebox{0.95\linewidth}{!}{
\begin{tabular}{c|c|ccccccccccc}
\toprule
\multicolumn{1}{l|}{}      & \#V  & \multicolumn{1}{c}{CVT} & \multicolumn{1}{c}{LpCVT} & \multicolumn{1}{c}{QEM} & \multicolumn{1}{c}{SMS} & \multicolumn{1}{c}{IEM} & \multicolumn{1}{c}{LPM} & \multicolumn{1}{c}{Ours} & \multicolumn{1}{c}{PQT} & \multicolumn{1}{c}{PQB} & \multicolumn{1}{c}{MD} & \multicolumn{1}{c}{ERB} \\ \midrule
\multirow{5}{*}{block}     & 0.5K   & 3.566/5.015            & 3.839/5.887              & 0.750                   & 33.397                  & 4.798                   & 112.789                 & 4.642/7.772              & 9.246                   & 9.112                  & 0.872                   & 1692.910            \\
                           & 1K     & 3.720/5.812             & 6.520/8.511              & 0.759                   & 33.818                  & 4.593                   & 112.163                 & 3.821/7.473              & 9.644                   & 9.784                   & 0.755                   & 1692.910            \\
                           & 3K     & 2.980/5.564             & 3.467/6.209              & 0.698                   & 32.482                  & 4.311                   & 118.774                 & 6.568/9.344              & 9.233                   & 9.739                   & 0.759                   & 1692.910            \\
                           & 5K     & 3.834/7.053           & 9.722/15.047             & 0.582                   & 33.172                  & 4.158                   & 118.615                 & 6.415/11.013              & 9.720                   & 9.845                   & 0.744                   & 1692.910            \\
                           & 10K    & 7.108/11.872           & 6.831/10.730             & 0.594                   & 28.692                  & 3.245                   & 134.092                 & 11.67/14.423              & 10.379                   & 10.760                   & 0.770                   & 1692.910            \\ \midrule
\multirow{5}{*}{bunny}     & 0.5K   & 2.417/3.681            & 2.347/3.618              & 0.346                   & 18.339                  & 2.897                   & 111.043                 & 2.963/7.477              & 5.762                   & 6.334                   & 0.536                   & 1542.080            \\
                           & 1K     & 3.622/5.722            & 3.79/5.203               & 0.476                   & 17.729                  & 2.929                   & 113.652                 & 5.085/7.301              & 5.936                   & 5.821                   & 0.580                   & 1542.080            \\
                           & 3K     & 6.839/12.335           & 2.836/5.672              & 0.392                   & 17.127                  & 2.664                   & 110.438                 & 6.896/12.300              & 5.489                   & 5.918                   & 0.508                   & 1542.080            \\
                           & 5K     & 6.516/13.238           & 2.569/5.724              & 0.449                   & 15.615                  & 2.350                   & 114.986                 & 7.587/13.183              & 6.203                   & 6.523                   & 0.541                   & 1542.080            \\
                           & 10K    & 5.324/16.906           & 6.829/14.073             & 0.377                   & 13.176                  & 2.275                   & 143.400                 & 10.588/19.440              & 6.839                   & 6.958                   & 0.489                   & 1542.080           \\ \bottomrule
\end{tabular}
}
\end{table*}

\subsection{Ablation Study}
\label{sec:ablation}
In the following, we present an ablation study regarding the weighting coefficients $\tau$, $\lambda_\text{NA}$,  $\lambda_\text{CVT}$ and the termination
tolerance $\mu$.

\begin{figure}[!h]
\centering
\begin{overpic}
[width=\linewidth]{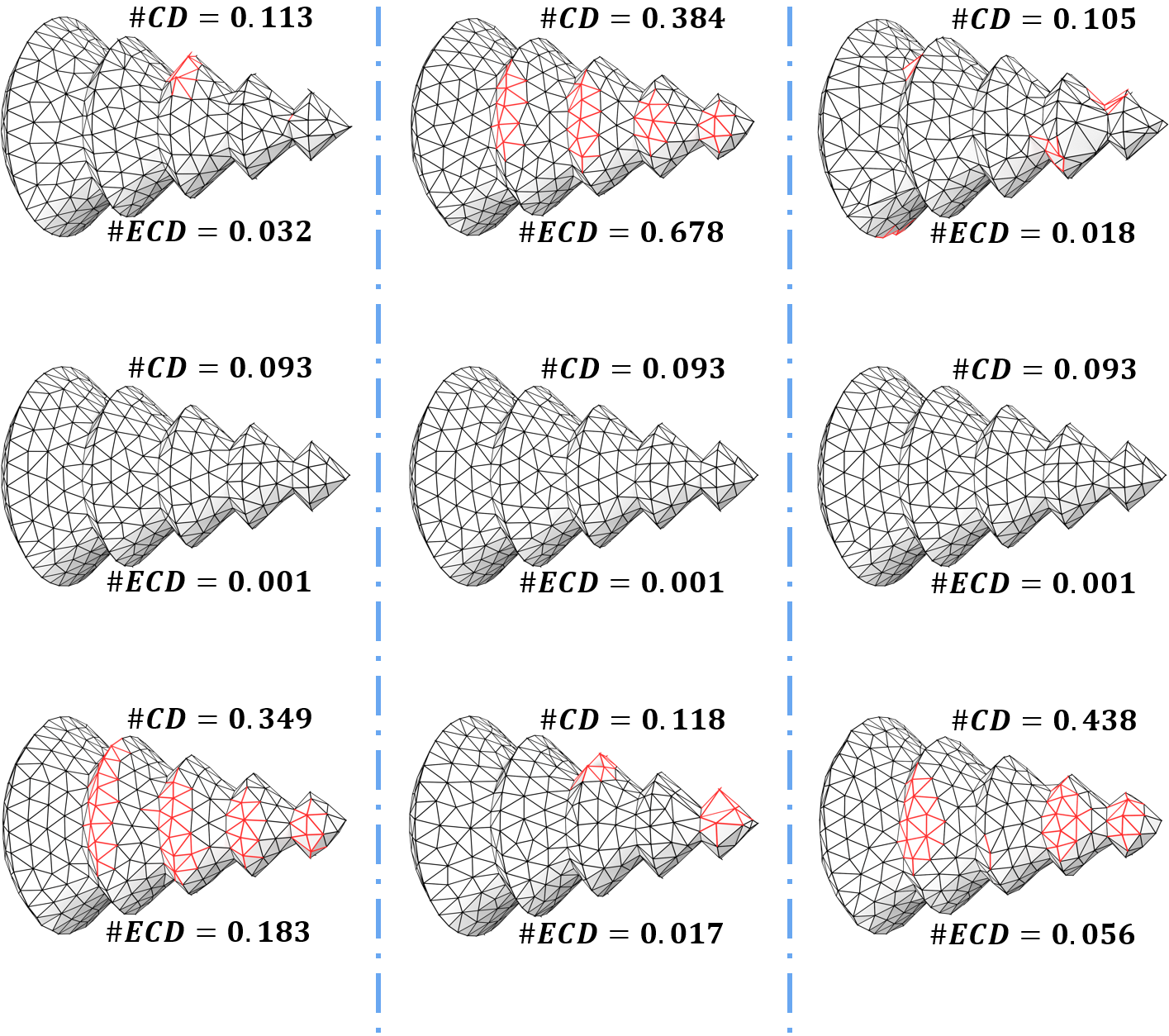}
\put(7, 63) {$\mathbf{\lambda_\text{CVT}=0.1}$}
\put(44,63) {$\mathbf{\lambda_\text{NA}=0.1}$}
\put(78, 63){$\mathbf{\tau=0.9}$}
\put( 3, 33) {$\mathbf{\lambda_\text{CVT}=1.0}$\textbf{(Df.)}}
\put( 40,33) {$\mathbf{\lambda_\text{NA}=1.0}$\textbf{(Df.)}}
\put(75, 33){$\mathbf{\tau=0.95}$\textbf{(Df.)}}
\put( 7, 4) {$\mathbf{\lambda_\text{CVT}=10.0}$}
\put(44, 4){$\mathbf{\lambda_\text{NA}=10.0}$}
\put(78, 4){$\mathbf{\tau=1.0}$}
\end{overpic}
\vspace{-3mm}
\caption{
Ablation study of the weighting coefficients $\tau$, $\lambda_\text{NA}$ and $\lambda_\text{CVT}$, where bad edges are colored in red.
We select
$\tau=0.95$, $\lambda_{\text{NA}}=1.0$, $\lambda_\text{CVT}=1.0$ as the favorite combination, which is used in all of our experiments.}
\label{fig:ablation}
\end{figure}
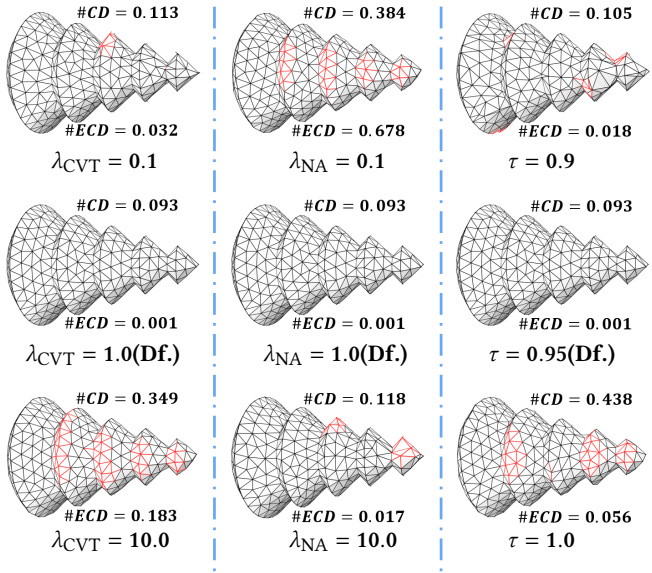

\begin{figure}[!h]
\centering
\begin{overpic}
[width=\linewidth]{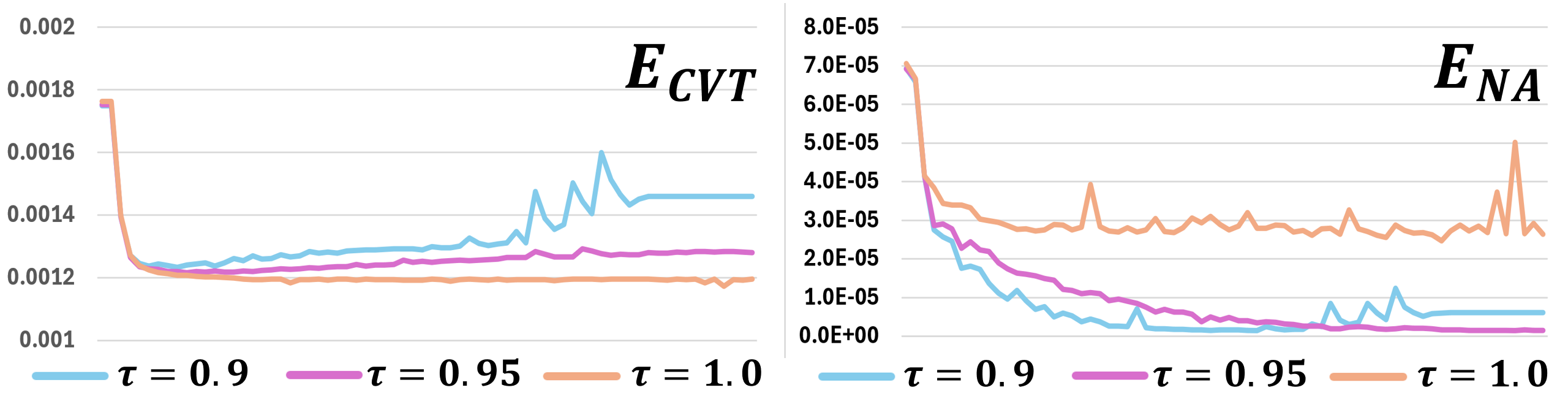}
\end{overpic}
\caption{
 \XR{
Plot curves about how the two energy terms $E_\text{CVT}$ and $E_\text{NA}$ decrease under various decay rates on the model depicted in Fig.~\ref{fig:ablation}.
}}
\label{fig:ablation3}
\end{figure}

\paragraph{Decaying Rate}
The decaying weight $\tau$ determines the rate of diminishing the CVT energy during the iteration process. When $\tau$ is set to $1.0$ (see the bottom-right part of Fig.~\ref{fig:ablation}), the optimization process excessively prioritizes the CVT term, leading to the degradation of some weak feature edges. If $\tau$ is slightly smaller than $1.0$, the contribution of the normal anisotropy term is increasingly emphasized with the growing number of iterations, while the CVT energy term is progressively suppressed. 
For example, by setting $\tau=0.95$, the weighting influence of the CVT energy reduces from $1.0$ to about $0.07$ after $50$ iterations. 
It can be imagined that if $\tau$ is too small (top right in Fig.~\ref{fig:ablation}), the point placement fails to achieve a good distribution since the CVT energy cannot sufficiently play its role before its influence disappears.
And when multiple feature lines are in close proximity, the algorithm necessitates uniformity to achieve an ideal outcome. In the absence of uniformity, the control region of a single Voronoi cell could encompass several feature lines, leading to the misalignment of features.
We show the numerical curves of the two energy terms $E_\text{CVT}$ and $E_\text{NA}$ under different decay rates in Fig.~\ref{fig:ablation3}.

\paragraph{Parameter $\lambda_\text{NA}$}
The parameter $\lambda_\text{NA}$ in the main paper defines the influence of the normal anisotropy term during optimization. As we increase $\lambda_\text{NA}$ from 0.1 to 10, the influence of the normal anisotropy term increases, and the meshing results resemble QEM.
When $\lambda_\text{NA}$ is set as large as $10.0$, the triangle quality degrades. Moreover, while accuracy appears to improve, this is not the case when the points are insufficient. See the red edges in the bottom-middle part of Fig.~\ref{fig:ablation}.



\paragraph{Parameter $\lambda_\text{CVT}$}
As Fig.~\ref{fig:ablation} illustrates, with the increasing value of $\lambda_\text{CVT}$ from $0.1$ to $1.0$, the contribution of the CVT energy term is enlarged, and the energy function gradually approaches the standard CVT. This makes the results resemble CVT, leading to an overall improvement in the triangulation quality of the model. However, it is observed that in the red region, the feature edges are damaged due to an excessive pursuit of triangulation quality.

\begin{figure}[!h]
\centering
\begin{overpic}
[width=\linewidth]{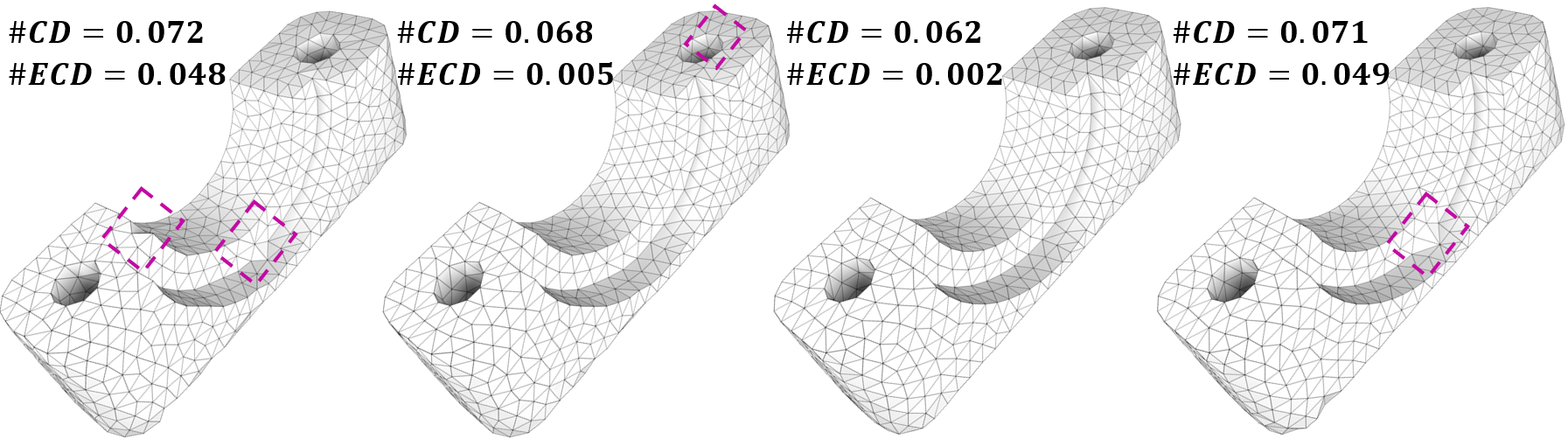}
\put( 7, -2) {$\mathbf{\mu=1.01}$}
\put(32, -2){$\mathbf{\mu=1.03}$}
\put(57, -2){$\mathbf{\mu=1.05}$}
\put(82, -2){$\mathbf{\mu=1.07}$}
\end{overpic}
\caption{
\XR{
Ablation study on the termination tolerance $\mathbf{\mu}$. We use $\mathbf{\mu=1.05}$ in all experiments.
}}
\label{fig:ablation2}
\end{figure}
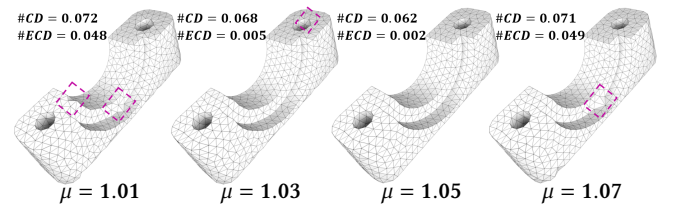

\paragraph{Parameter $\mathbf{\mu}$}
\XR{
We present the ablation study on the termination tolerance $\mathbf{\mu}$ in Fig.~\ref{fig:ablation2}. It can be observed that a smaller $\mathbf{\mu}$, such as 1.01 or 1.03, requires fewer iterations to terminate, but the feature alignment ability is not fully realized. Conversely, a larger $\mathbf{\mu}$ value, such as 1.07, requires hundreds of iterations to reach termination. Furthermore, the triangle quality may deteriorate at termination. This is why we use $\mathbf{\mu=1.05}$ in all experiments.
}



\subsection{Non-Euclidean Metric}
\label{sec:nonE}
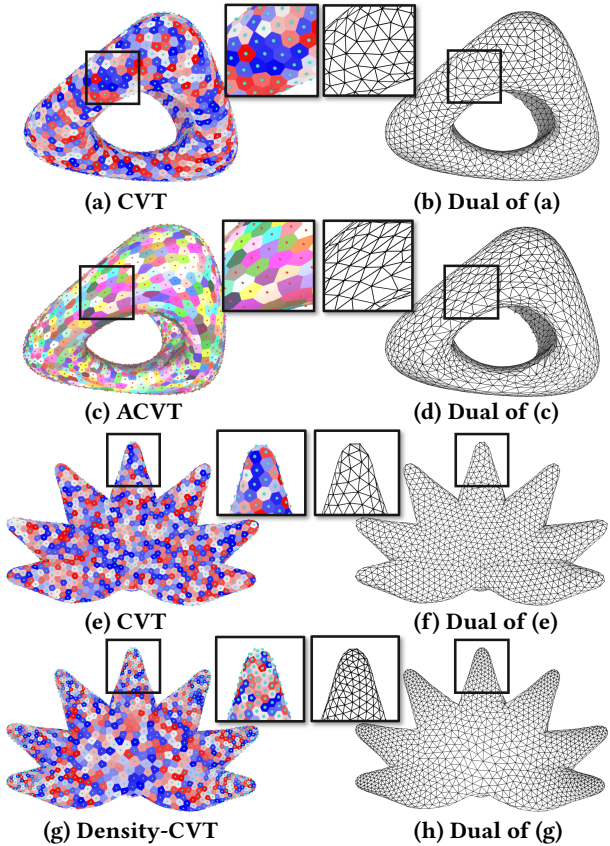
\begin{figure}[!h]
    \centering
    \begin{overpic}[width=0.95\linewidth]{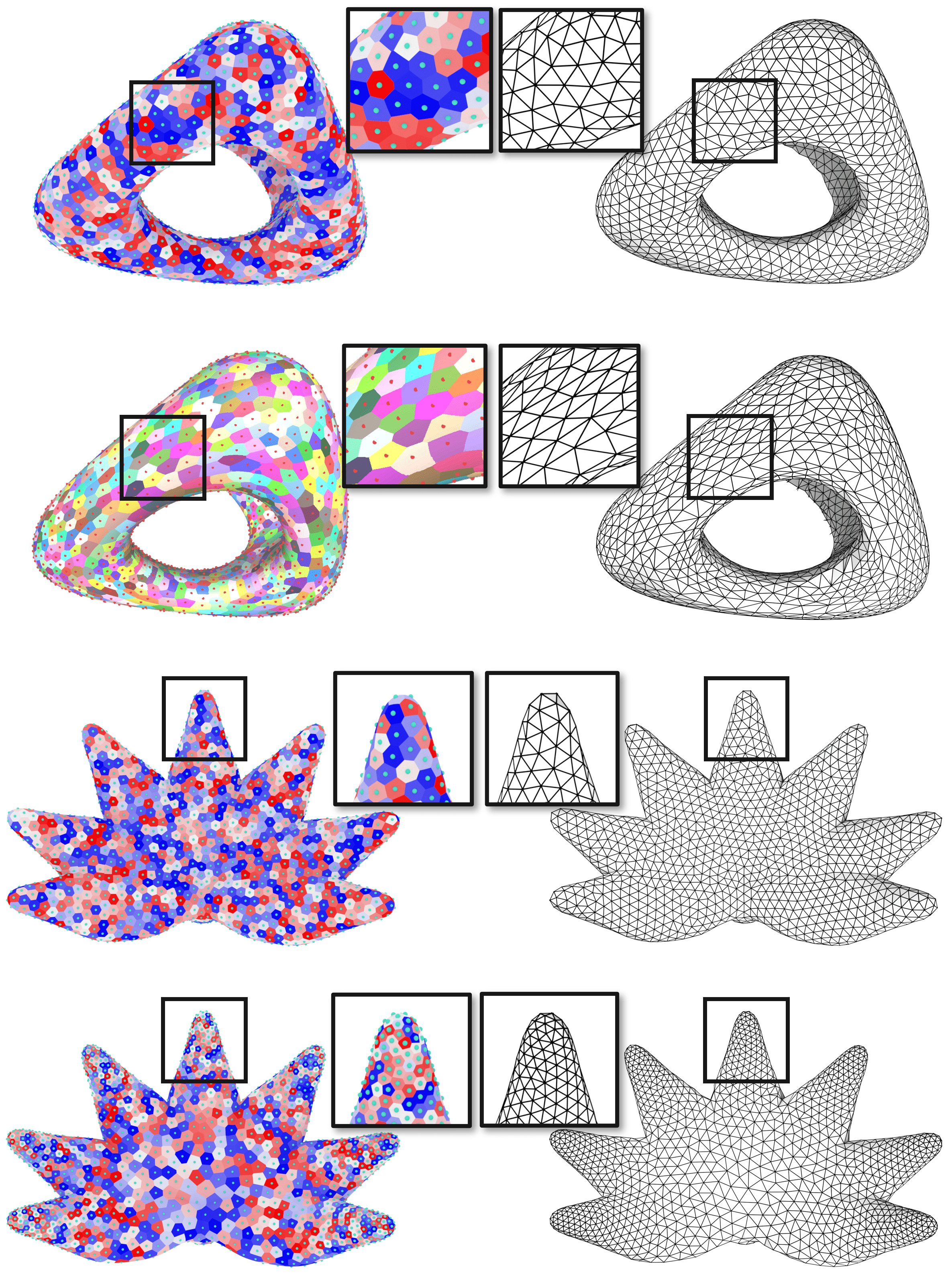}

    \put(10, 74.5){\textbf{(a) CVT}}
    \put(50, 74.5){\textbf{(b) Dual of (a)}}
    \put(10, 49){\textbf{(c) ACVT}}
    \put(50, 49){\textbf{(d) Dual of (c)}}
    \put(10, 23.5){\textbf{(e) CVT}}
    \put(50, 23.5){\textbf{(f) Dual of (e)}}
    \put(5, -2){\textbf{(g) Density-CVT}}
    \put(50, -2){\textbf{(h) Dual of (g)}}
    \end{overpic}
    \vspace{5pt}
    \caption{Examples of non-Euclidean distance, (a-d) is the anisotropic metrics and RVDs, (e-h) is the density-based metrics and RVDs.}
    \label{fig:anisotropic}
\end{figure}

It is worth noting that we use Euclidean metric-based RVD in our experiments, and we can also support non-Euclidean metrics. 
\paragraph{Anisotropic Metric}
As Eq.~(\ref{eq:M}) demonstrates, our formulation supports non-Euclidean metrics. For instance, one can utilize curvature to define an anisotropic field on the surface, establishing curvature-aware point-to-point distances. An example is illustrated in Fig.~\ref{fig:anisotropic}, where the surface decomposition is computed based on plane cutting~\cite{sun2011obtuse}. 
The clipping technique, based on our tests, can only handle convex shapes (e.g., an ellipsoid). To the best of our knowledge, there is currently no mature solver available for the anisotropic Restricted Voronoi Diagrams.

\paragraph{Density Function}
It's natural that our algorithm also supports density-adaptive remeshing. 
The typical density field is derived from the local feature size~(LFS), adapting the triangle sizes to the curvature. As depicted in Fig.~\ref{fig:anisotropic}, the utilization of density enables the concentration of mesh elements around features.

\subsection{Robustness and Scalability}
\label{sec:comparison_varying}
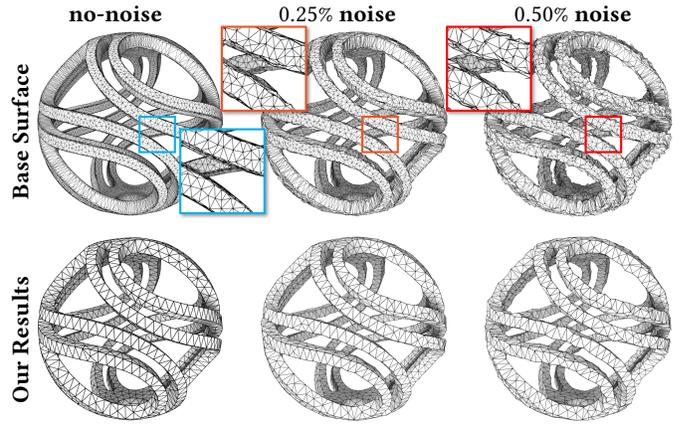
\begin{figure}[!h]
\centering
\vspace{8pt}
\begin{overpic}
[width=0.99\linewidth]{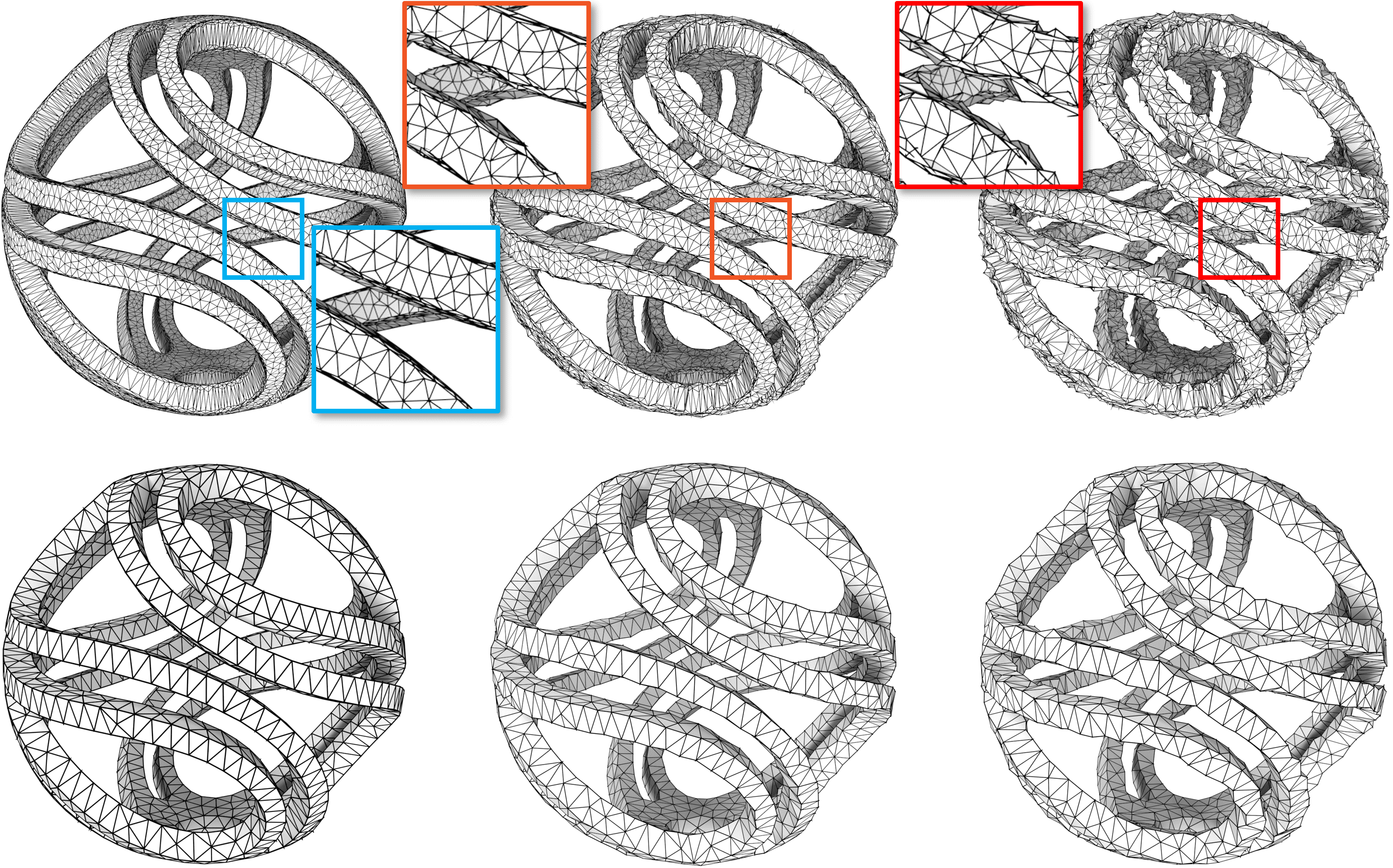}
\put( 5, 63) {\textbf{{no-noise}}}
\put(38, 63) {\textbf{{$0.25\%$ noise}}}
\put(75, 63) {\textbf{{$0.50\%$ noise}}}
\put(-4, 3) {\rotatebox{90}{\textbf{Our Results}}}
\put(-4, 35){\rotatebox{90}{\textbf{Base Surface}}}
\end{overpic}
\caption{
Ablation study of our noise-resistant ability. We add $0.25\%$ and $0.50\%$ Gaussian noise to the surface vertices of the model. More noises may results in bad surface connectives such as self-intersections. Our method shows strong resistance to noise to some extent. }
\label{fig:noise}
\vspace{-5pt}
\end{figure}

\paragraph{Noise}
\label{sec:ablation_noise}
To test our noise-resistant ability, we add $0.25\%$ and $0.50\%$ Gaussian noise to the surface vertices of the \textit{curve-ball} model, as shown in Fig.~\ref{fig:noise}. As the noise level increases, the triangle quality of the mesh is diminished, and the surface becomes very bumpy. It can be observed from Fig.~\ref{fig:noise} that our method exhibits strong resistance to noise, consistently producing robust outputs.

\begin{figure}[!h]
\centering
\begin{overpic}
[width=0.99\linewidth]{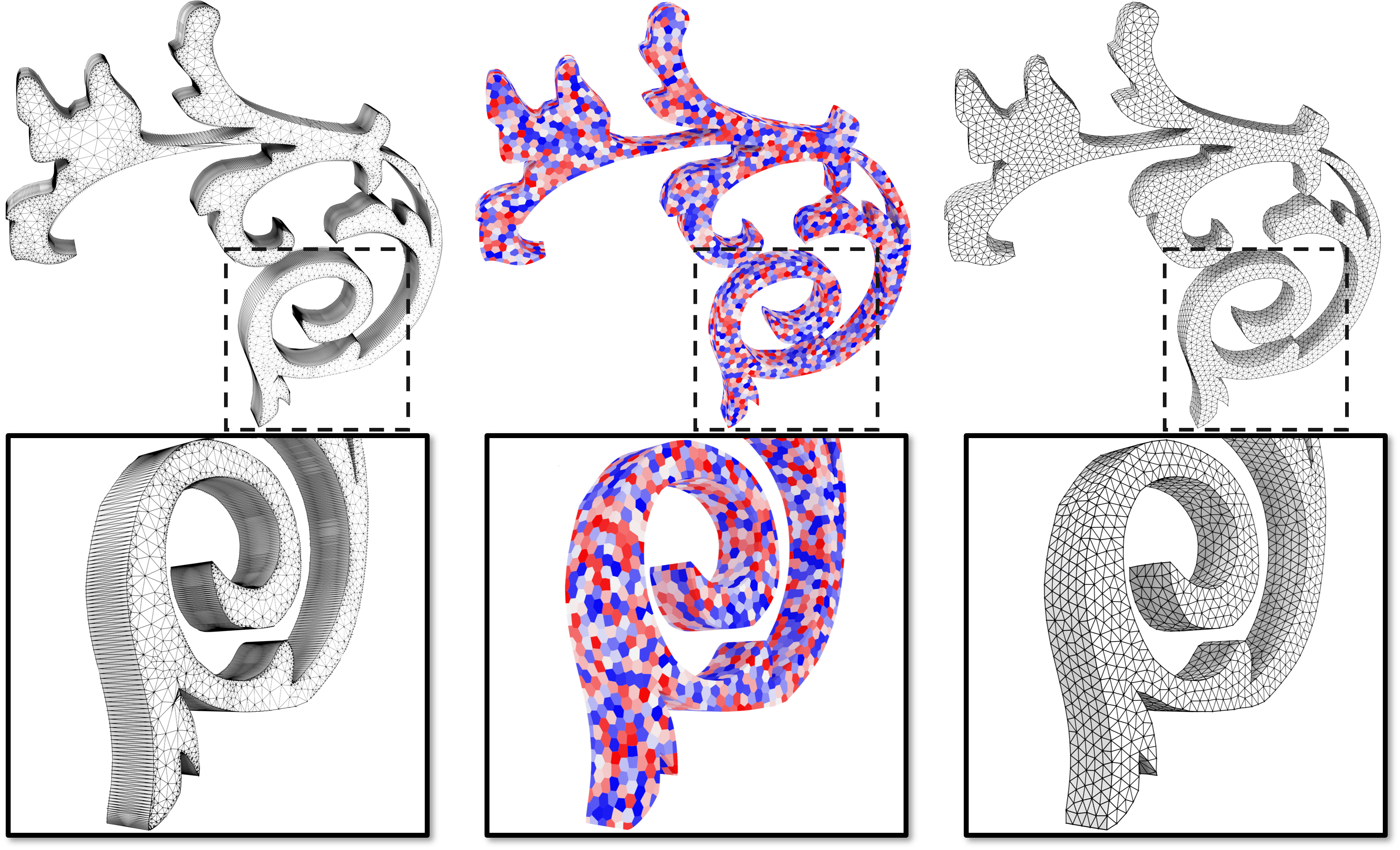}
\put(5, -3) {\textbf{Base Surface}}
\put(46, -3){\textbf{RVD}}
\put(79, -3){\textbf{Result}}

\end{overpic}
\caption{
Ablation study on the low triangle quality of the base surface. The input mesh contains thin and sharp triangles, which are normally undesired. Our method demonstrates resistance to these low-quality meshes as input and outputs a high-quality simplification result.
}
\label{fig:lowTri}
\vspace{-5pt}
\end{figure}

\paragraph{Pooly Triangulated Inputs.}
\label{sec:ablation_base_quality}
Our algorithm optimizes a set of movable points on the surface. It repeatedly decomposes the surface into regions during optimization. Therefore, it depends only on the geometry and works well for poorly triangulated inputs. 
Fig.~\ref{fig:lowTri} validate the robustness of our algorithm.
We conducted tests on the \textit{artwork} model, which has poor mesh quality on its side. 

\paragraph{Initialization}

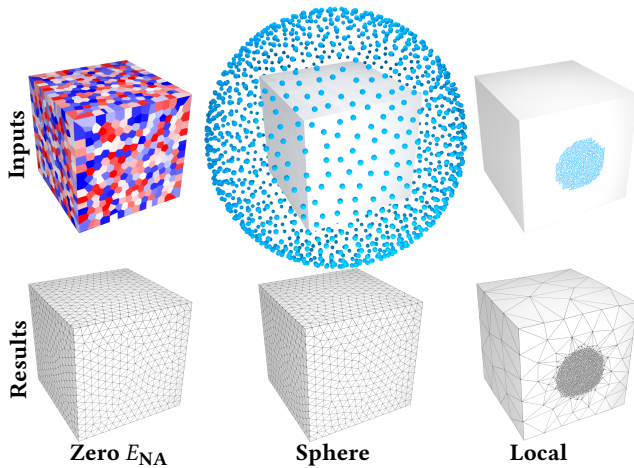
\begin{figure}[!h]
\centering
\begin{overpic}
[width=0.95\linewidth]{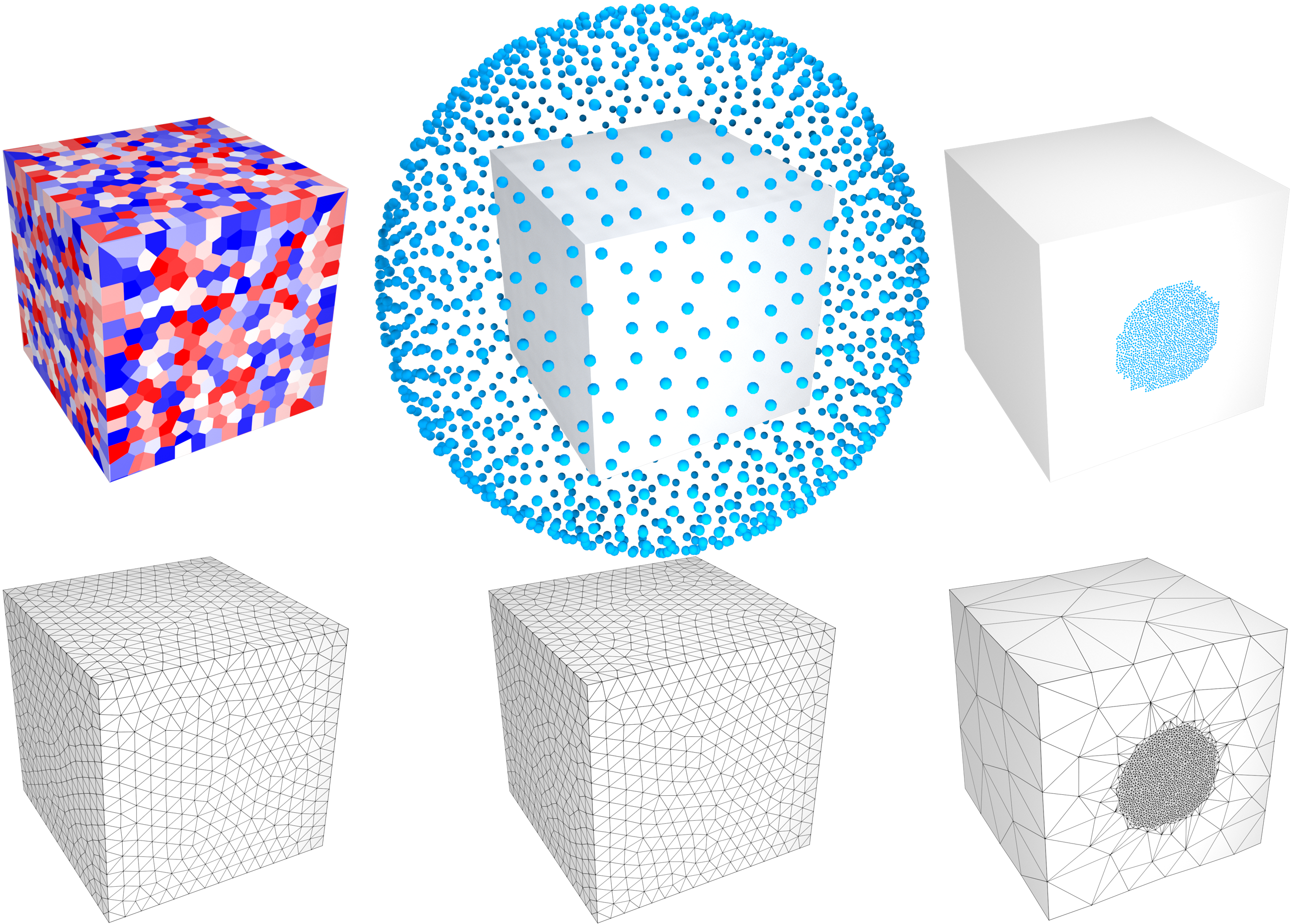}
\put(-3, 43){\rotatebox{90}{\textbf{Inputs}}}
\put(-3, 7){\rotatebox{90}{\textbf{Results}}}
\put( 7, -3){\textbf{Zero $E_\text{NA}$}}
\put(44, -3){\textbf{Sphere}}
\put(79, -3){\textbf{Local}}
\end{overpic}
\caption{
\XR{Our results with different point placement initialization strategies.}}
\label{fig:Init}
\end{figure}

\XR{
The ``normal anisotropy''  term can also reach its minimum when the Voronoi cell edges align with the feature lines of the model, which is an unstable state, as mentioned in Section~\ref{subsec:Links}. We construct such an unstable example in the first column of Fig.~\ref{fig:Init}. Our method can still optimize it into a stable point placement.

Additionally, we explore two additional point placement initialization strategies. The first assumes the points are distributed on a spherical surface, and the second assumes the points gather around a single point. In the second case, it can be seen that the points encounter certain resistance when crossing a feature line. Therefore, we recommend the Poisson-disk sampling strategy for initialization in all our experiments.
}

\begin{figure}[!h]
\centering
\begin{overpic}
[width=0.99\linewidth]{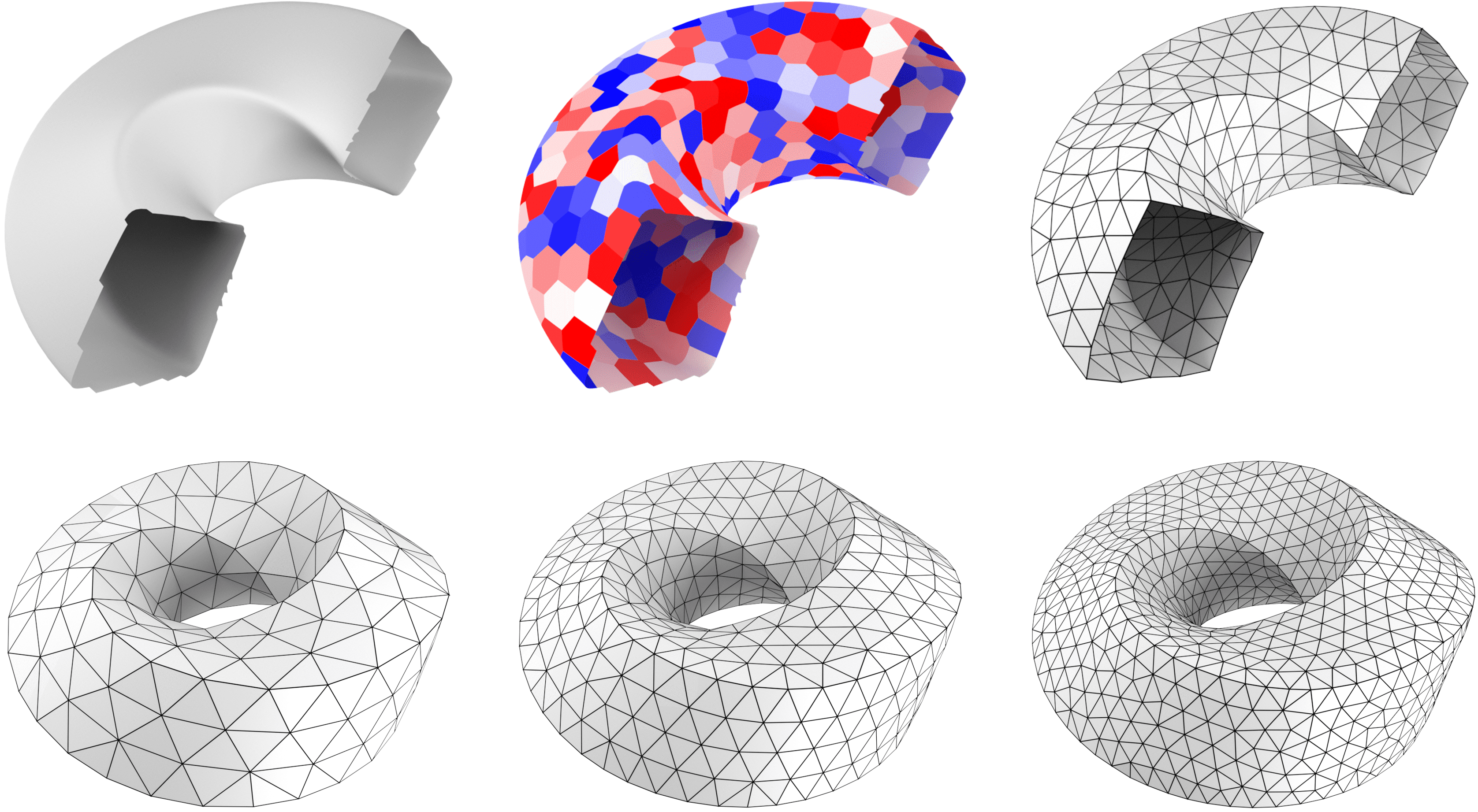}
\put(4, 25) {\textbf{Open Surface}}
\put(44, 25){\textbf{Our RVD}}
\put(75, 25){\textbf{Our Result}}
\put(7, -3) {$\mathbf{\#V=200}$}
\put(44, -3){$\mathbf{\#V=500}$}
\put(75, -3){$\mathbf{\#V=1000}$}
\end{overpic}
\caption{
Ablation study on an open surface (top) and three different target numbers of vertices, $\#V$, (bottom).
}
\label{fig:open_sampling}
\vspace{-5pt}
\end{figure}
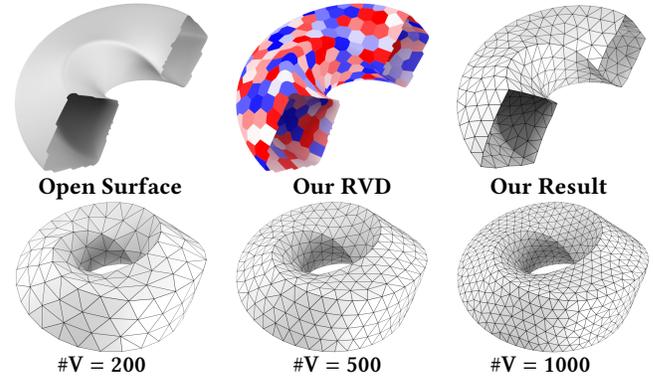

\paragraph{More Tests.}
\label{sec:ablation_open_sampling}
Generally, our method takes a closed surface as the input. However, it must be noted that our method can also be applied to meshes with open boundaries, as shown in the top of Fig.~\ref{fig:open_sampling}. In addition, users can specify the target number of vertices. Despite the disparity in meshing resolution ($200$, $500$, and $1000$), our algorithm can consistently meet the requirements of accuracy, triangle quality, and feature alignment; see the bottom of Fig.~\ref{fig:open_sampling}.

\subsection{Potential Applications}

\begin{figure}[!tp]
\centering
\begin{overpic}
[width=0.97\linewidth]{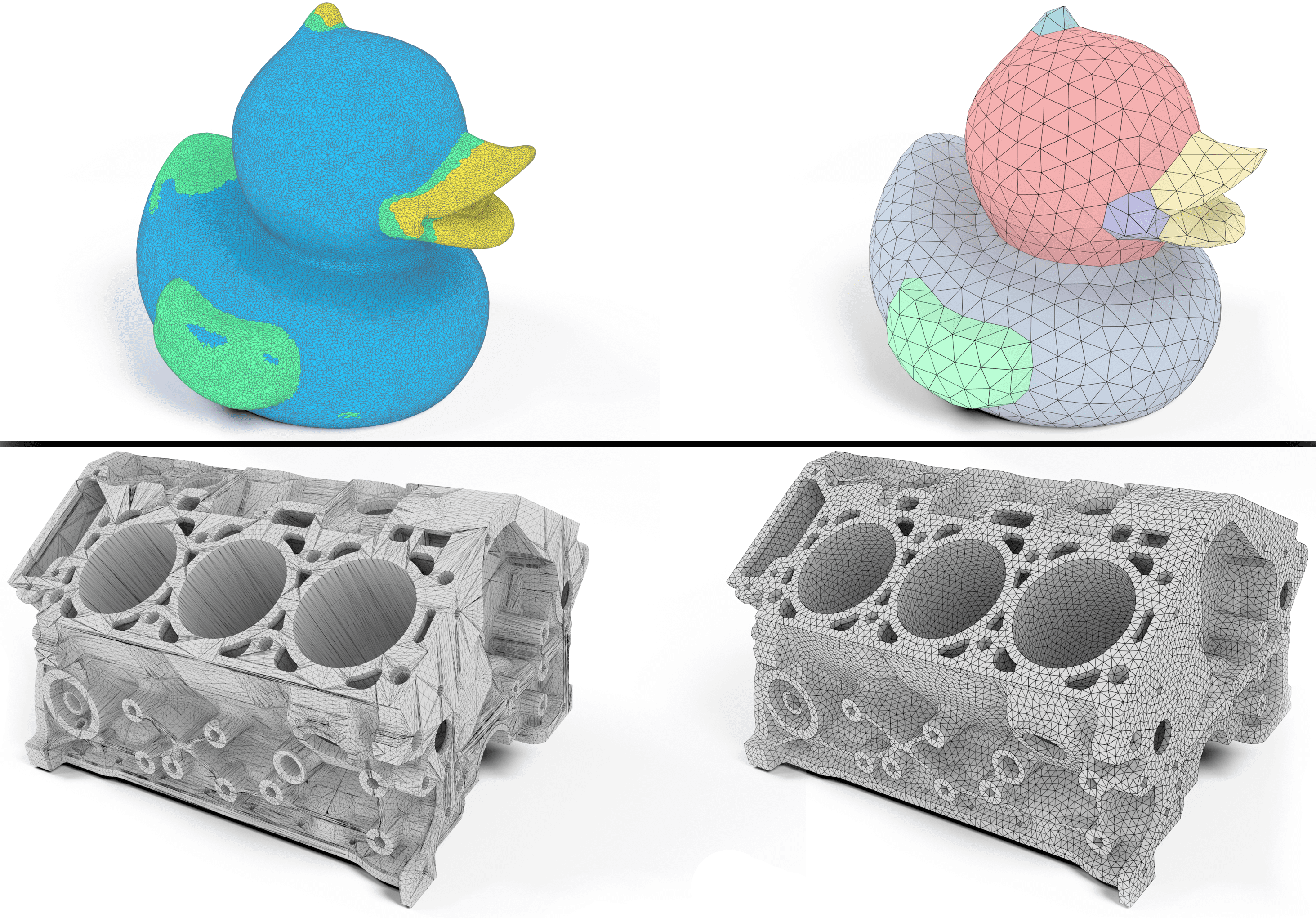}
\put( 7, 40){\textbf{(a)}}
\put(62, 40){\textbf{(b)}}
\put(7, 3){\textbf{(c)}}
\put(62, 3){\textbf{(d)}}
\put(-3,5){\rotatebox{90}{\textbf{Lightweighting}}}
\put(-3,40){\rotatebox{90}{\textbf{Segmentation}}}
\end{overpic}
\vspace{-3mm}
\caption{
\XR{Two typical applications on mesh segmentation and CAD model lightweight representation.}}
\label{fig:application}
\end{figure}

\paragraph{Mesh Segmentation}
\XR{
Past research indicates that feature lines are crucial for mesh segmentation. However, existing segmentation algorithms struggle to align with weak feature lines. It can be imagined that by simplifying the mesh while preserving its weak features, segmentation becomes more manageable.

Using~\cite{shapira2008consistent} as the segmentation solver, it is evident from Fig.~\ref{fig:application}~(a,b) that the segmentation results are significantly improved. This enhancement is due to our algorithm's ability to consolidate weak features during mesh simplification and reduce the impact of intricate details.

In summary, our algorithm can serve as a preprocessing step for the mesh segmentation task. It not only eases the difficulty of mesh segmentation but also produces visually straight segmentation boundaries.

}

\paragraph{Lightweighting}
\XR{
In the realm of Computer-Aided Design (CAD), mesh simplification is a vital tool for achieving model lightweighting—a process that reduces the complexity of CAD models without compromising their functional integrity or design intent. Simplification is crucial for managing intricate CAD models that can be computationally demanding due to their high polygon counts. 
As seen in Fig.~\ref{fig:application} (c), the original CAD model is displayed with all its detailed features, which, while accurate, results in a heavy model that requires significant computational resources. 
In contrast, Fig.~\ref{fig:application} (d) showcases the CAD model post-simplification, where the mesh has been effectively reduced in complexity. This streamlined version preserves the model's fundamental geometry and essential characteristics, yet is lighter in terms of data size and easier to manipulate and render. The lightweight model offers numerous benefits, including faster loading times, enhanced performance in real-time visualization, and more efficient storage and transfer. Additionally, it supports more effective collaboration and integration with other systems (e.g., VR devices) that have lower hardware specifications.
}

\begin{figure}[!tp]
\centering
\begin{overpic}
[width=0.85\linewidth]{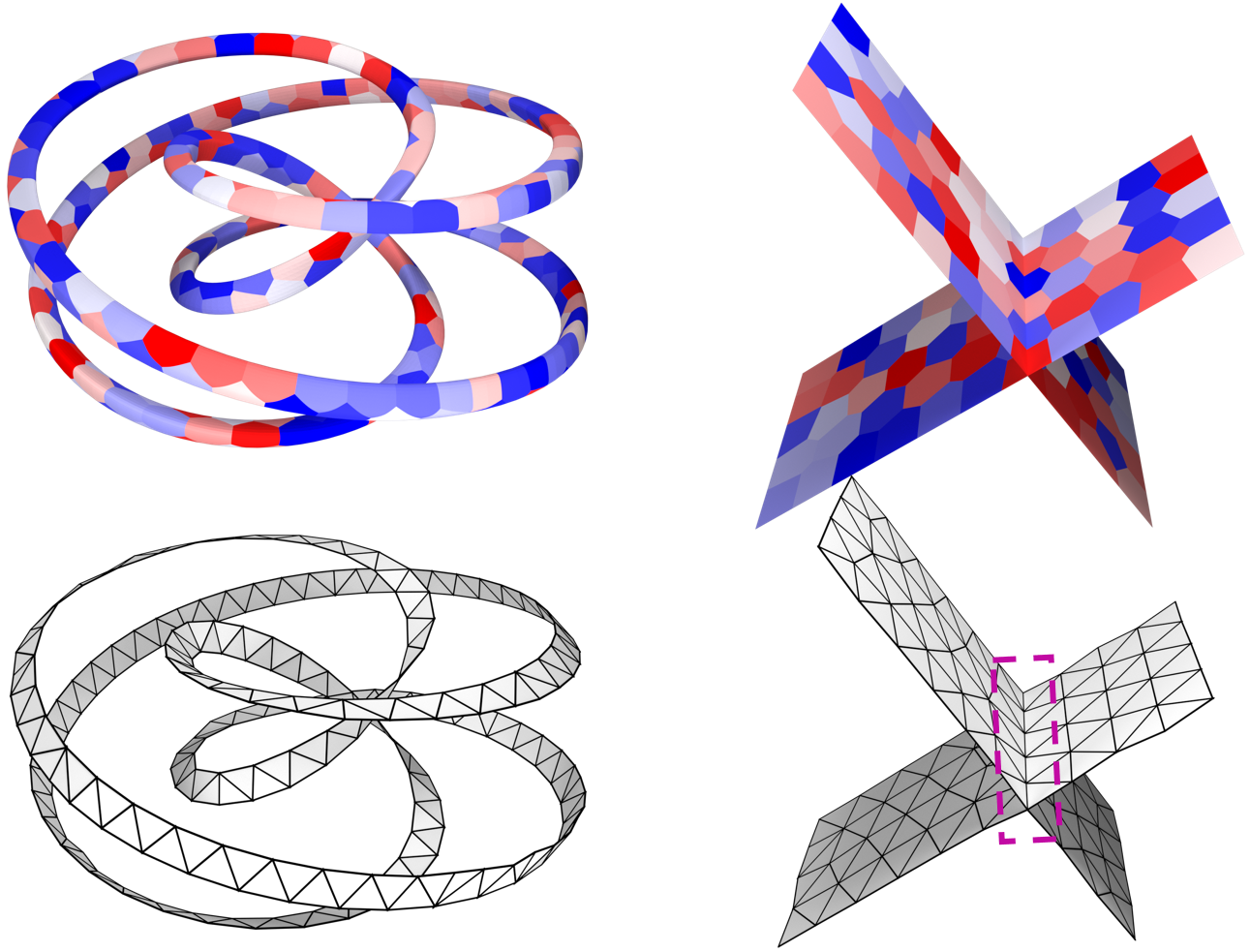}
\put(14, -3){\textbf{Thin tubes}}
\put(61,-3){\textbf{Self-intersection}}
\end{overpic}
\caption{
\XR{Our algorithm, in its current form, cannot handle cases where the target number of points is too few. Additionally, it struggles with input models that have self-intersections.}}
\label{fig:limitation}
\end{figure}

\subsection{Limitations}


\XR{
Our methodology has some limitations. The first arises from the lack of a robust solver capable of computing anisotropic Restricted Voronoi Diagrams (RVDs). Consequently, our approach does not fully exploit the potential to produce anisotropic meshes that align with features of interest. The second limitation relates to the suboptimal efficiency of our implementation, which could benefit from GPU-based acceleration.
The third limitation, but certainly not the least, is that the RVD strategy proposed in this paper, while simple and effective, may fail when the target number of points is too few. This is because the dual of the RVD cannot define a manifold triangle mesh, as demonstrated in the first example in Fig.~\ref{fig:limitation}. However, it is also noted that our RVD strategy performs well as long as the target number of points is sufficient, based on our numerous tests.
We also present a scenario where two distinct faces intersect; our technique will lead to the occurrence of a non-manifold artifact. This implies that when the input has self-intersections, our algorithm may produce non-manifold vertices/edges.


}



\section{Conclusions}
In this paper, we propose an objective function that concurrently integrates the requirements of accuracy, triangle quality, and feature alignment. Our function includes the normal anisotropy term and the CVT energy term, balanced with a decaying weight. We conducted extensive experiments to compare our approach with existing state-of-the-art (SOTA) methods, validating its efficacy. Notably, our approach not only meets multiple requirements but also consolidates weak features, setting it apart from other SOTA methods. Additionally, we introduce a simple yet effective technique for computing RVDs on thin-plate models.



\begin{acks}
The authors would like to thank the anonymous reviewers for their valuable comments and suggestions. This work is supported by National Key R\&D Program of China (2022YFB3303200), National Natural Science Foundation of China (62272277, U23A20312, 62072284) and NSF of Shandong Province (ZR2020MF036). 
Ningna Wang and Xiaohu Guo were partially supported by National Science Foundation (OAC-2007661).
Zichun Zhong was partially supported by National Science Foundation (OAC-1845962, OAC-1910469, and OAC-2311245).
\end{acks}

\FloatBarrier
\bibliographystyle{ACM-Reference-Format}
\bibliography{sample-base}



\end{document}